%% file: KKDS19.tex
\documentclass[smallcondensed,natbib]{svjour3}
\smartqed

\usepackage[leftcaption]{sidecap}
\usepackage{graphicx}
\usepackage{mathptmx}
\usepackage{mdframed}

\journalname{Empirical Software Engineering}

\usepackage{etoolbox}
\apptocmd{\sloppy}{\hbadness 10000\relax}{}{}
\apptocmd{\sloppy}{\vbadness 10000\relax}{}{}
\usepackage{array}
\usepackage{booktabs}
\usepackage{verbatim}
\usepackage{amsmath,amssymb,amsfonts}
\usepackage{algorithmic}
\usepackage[export]{adjustbox}
\DeclareGraphicsExtensions{.pdf,.jpeg,.png}
\usepackage{textcomp}
\usepackage{xcolor}
\usepackage{url}
\usepackage[noadjust]{cite}
\def\BibTeX{{\rm B\kern-.05em{\sc i\kern-.025em b}\kern-.08em
    T\kern-.1667em\lower.7ex\hbox{E}\kern-.125emX}}
\newcolumntype{L}[1]{>{\raggedright\let\newline\\\arraybackslash\hspace{0pt}}m{#1}}
\newcommand{\msr}{\textsc{msr}}

\newmdtheoremenv[%
  outerlinewidth=4pt,
  leftmargin=4em,
  rightmargin=4em,
  innertopmargin=4pt,
  splittopskip=\topskip,
  skipbelow=\baselineskip,
  skipabove=.5\baselineskip]{implication}{Implication}

\usepackage{draftwatermark}
\SetWatermarkText{\textsc{Preprint}}
\SetWatermarkScale{.5}
\SetWatermarkColor[gray]{.83}

\usepackage{tikz}
\usepackage{lipsum}
\newcommand\copyrighttext{%
  \footnotesize \textcopyright This is a post-peer-review, pre-copyedit version of an article published in Empirical Software Engineering journal. The final authenticated version is available online at: https://doi.org/10.1007/s10664-020-09834-7.}
\newcommand\copyrightnotice{%
\begin{tikzpicture}[remember picture,overlay]
\node[anchor=south,yshift=150pt,xshift=-52pt] at (current page.south) {\fbox{\parbox{\dimexpr\textwidth-\fboxsep-\fboxrule\relax}{\copyrighttext}}};
\end{tikzpicture}%
}

\begin{document}

\title{Standing on Shoulders or Feet? \\
An Extended Study on the Usage of the MSR Data Papers
}

\author{Zoe Kotti   \and
        Konstantinos Kravvaritis    \and \\
        Konstantina Dritsa  \and
        Diomidis Spinellis
}

\institute{
    Zoe Kotti, Konstantinos Kravvaritis,
    Konstantina Dritsa, Diomidis Spinellis   
    \at{Department of Management Science and Technology \\
    Athens University of Economics and Business} \\
    \email{\{zoekotti, kravvaritisk, dritsakon, dds\}@aueb.gr} \\
    ORCID: \{0000-0003-3816-9162, 0000-0002-8889-0612,
    0000-0003-3395-2182, 0000-0003-4231-1897\}     
}

\date{}

\maketitle
\copyrightnotice

\begin{abstract}

\emph{Context:}
The establishment of the Mining Software Repositories (MSR)
data showcase conference track has encouraged researchers
to provide data sets as a basis for further empirical studies. \\
\emph{Objective:}
Examine the usage of data papers published
in the MSR proceedings in terms of use frequency,
users, and use purpose. \\
\emph{Method:}
Data track papers were collected from the MSR data showcase track
and through the manual inspection of older MSR proceedings.
The use of data papers was established through manual citation searching
followed by reading the citing studies
and dividing them into strong and weak citations.
Contrary to weak,
strong citations truly use the data set of a data paper.
Data papers were then manually clustered based on their content,
whereas their strong citations were classified by hand
according to the knowledge areas
of the Guide to the Software Engineering Body of Knowledge.
A survey study on 108 authors and users of data papers
provided further insights
regarding motivation and effort in data paper production,
encouraging and discouraging factors in data set use,
and future desired direction regarding data papers. \\
\emph{Results:}
We found that 65\% of the data papers have been used in other studies,
with a long-tail distribution in the number of strong citations.
Weak citations to data papers usually refer to them
as an example.
MSR data papers are cited in total less than other MSR papers.
A considerable number of the strong citations stem
from the teams that authored the data papers.
Publications providing Version Control System (VCS) primary and derived data are the most frequent
data papers and the most often strongly cited ones.
Enhanced developer data papers are the least common ones, and the
second least frequently strongly cited.
Data paper authors tend to gather data in the context of other research.
Users of data sets appreciate high data quality
and are discouraged by lack of replicability of data set construction.
Data related to machine learning or derived from the manufacturing sector
are two suggestions of the respondents for future data papers. \\
\emph{Conclusions:}
Data papers have provided the foundation for a significant number of studies,
but there is room for improvement in their utilization.
This can be done by
setting a higher bar for their publication,
by encouraging their use, 
by promoting open science initiatives, and
by providing incentives for the enrichment of existing data collections.


\end{abstract}

\begin{quote}
\emph{
``Indeed, one of my major complaints about the computer field
is that whereas Newton could say,
`If I have seen a little farther than others,
it is because I have stood on the shoulders of giants,'
I am forced to say, `Today we stand on each other's feet.'
Perhaps the central problem we face in all of computer science
is how we are to get to the situation
where we build on top of the work of others
rather than redoing so much of it in a trivially different way.''
}
\end{quote}
\hfill --- Richard Wesley Hamming\footnote{1968 ACM Turing Award Lecture~\citep{Ham69}}

\section{Introduction}
\label{sec:intro}

Software engineering data sets are often a key ingredient for
performing empirical software engineering by testing a hypothesis
through an experiment run on such data~\citep{Cuk05}.
They can be used to empirically evaluate
software product quality
and development process attributes
and also to create or verify estimation models~\citep{LS12}.
In addition, publicly available data sets can help researchers perform
so-called \emph{exact} replications of existing
studies and thus address potential internal validity problems~\citep{SCVJ08}.
These, in contrast to \emph{conceptual} replications, which
follow an independently developed experimental procedure,
attempt to control as many factors of the original study as
possible, varying almost no (in \emph{dependent} replications) or
only some (in \emph{independent} replications) conditions of the
experiment~\citep{SCVJ08}.

Yet, at least in the past, data sets for software engineering research
were small in size and difficult to obtain~\citep{KPPJ02}.
The situation has improved over the past decades
with the emergence of open source software~\citep{KH06},
and the growing interest in sharing artifacts, including data sets,
along with research publications.
Such efforts are encouraged by initiatives such as
the \textsc{acm sigsoft} Artifact Evaluation Working Group,\footnote{\url{https://github.com/acmsigsoft/artifact-evaluation}}
which aims to integrate the artifact evaluation in the publication process,
or the \emph{Recognizing and Rewarding Open Science in Software Engineering} (\textsc{rose}) festival,
which salutes replication and reproducibility in software engineering.
For these reasons researchers have collaborated~\citep{Cuk05}
through various initiatives to develop data set repositories,
such as the \emph{International Conference on Predictive Models and
Data Analytics for Software Engineering} (\textsc{promise})~\citep{SM05},
or to promote the sharing and publication of data, as through
the \textsc{us} National Institute of Standards and Technology's
``Error, Fault, and Failure Data Collection and Analysis Project''~\citep{Wal98},
the Mining Software Repositories (\msr) conference data showcase track~\citep{MSR13},
or the \emph{awesome-msr}
GitHub project.\footnote{\url{https://github.com/dspinellis/awesome-msr}}

The \msr\ data showcase track, established in 2013,
aims at encouraging the research community to develop, share, and document
software engineering research data sets.
In the words of the 2013 \msr\ conference chairs~\citep{ZMSD13}:
\begin{quote}
``rather than describing research achievements, data papers describe
datasets curated by their authors and made available to others. Such
papers provide description of the data, including its source; methodology
used to gather it; description of the schema used to store it, and any
limitations and/or challenges of this data set.''
\end{quote}

In the past decade tens of data set papers have been published in the
\msr\ conference.
Given the effort that went into creating the data sets and publishing the
corresponding papers, it is reasonable to investigate what the outcome
has been.
This study aims to answer the question by examining
the usage of the data papers published
in the \msr\ proceedings in terms of
use frequency --- to evaluate the data track's actual impact,
users --- to examine researchers' potential reluctance
to work with data coming outside their organization,
and use purpose --- to identify the most used types of data papers,
and the types of studies that mainly use them.
The study's contributions are:
\begin{itemize}
\item the systematic collection of research that has been based on
\msr\ data papers,
\item the categorization of the subjects tackled using \msr\ data papers,
\item the quantitative analysis of the \msr\ data papers' impact, and
\item the analysis of the community's opinion regarding data paper publication and use.
\end{itemize}
In the following Section~\ref{sec:related} we present an overview
of related work.
We then describe our study's methods in Section~\ref{sec:methods},
present our results in Section~\ref{sec:results},
discuss the findings, and identify our study's implications
in Section~\ref{sec:discussion}.
The study is complemented
by the associated validity threats in Section~\ref{sec:validity},
followed by our conclusions in Section~\ref{sec:conclusions}.
The data sets associated with our study
(data papers, citing papers, categorizations, \msr\ papers,
citations, survey questionnaire and responses) are made available online.\footnote{\url{http://doi.org/10.5281/zenodo.3709219}}

A shorter version of this study appeared
in the 2019 \textsc{ieee}/\textsc{acm}
16th International Conference on Mining Software Repositories
(\msr\ '19)~\citep{KS19}.
This work extends the conference paper by
using multiple raters and established research methods
for the manual clustering of data papers
and the classification of strong citations.
Furthermore,
this work introduces a questionnaire survey study
on all identified primary authors and users of data papers,
and an analysis of weak citations
(defined in Section~\ref{sec:citation-methods})
to data papers.

\section{Related Work}
\label{sec:related}

A variety of evaluations have been conducted
through research analysis.
We recognize two major fields of evaluations:
surveys and bibliometrics.
Surveys review and summarize previously published studies
of a particular topic through qualitative analysis.
Webster and Watson~\citeyearpar{WW02} have authored a thorough guide
on writing high quality literature reviews.
On the other hand, bibliometric studies are statistical analyses
of publication data.
We consider our work part of the bibliometric research strand,
and to the best of our knowledge, we are the first
to conduct a quantitative review of data paper usage.

A first step toward the assistance of bibliometric research
in the field of software engineering models was made in
2004 by the organizers of the \textsc{promise} workshop,
in their attempt ``to strengthen the community's faith
in software engineering models''~\citep{Cuk05}.
Authors of such models were asked to submit, along with their work,
a related data set to the \textsc{promise} repository.

Many individuals have carried out
interesting quantitative bibliometric research on various topics.
Robles~\citeyearpar{Rob10} conducted bibliometric research on papers
that contained experimental analyses of software projects
and were published in the \msr\ proceedings
from 2004--2009.
His objective was to review their potential replicability.
The outcome proves that \msr\ authors prefer
publicly available data sources from free software repositories.
However, the amount of publicly available processed data collections
was very low at the time, a fact we also state in Section~\ref{sec:rq1}.
Concerning replicability, Robles found that only a limited number
of publications are replication friendly.

Liebchen and Shepperd~\citeyearpar{LS08} performed
a different quantitative analysis on data sets.
Their aim was to assess quality management techniques
used by authors when producing data collections.
They found that a surprisingly small percentage of studies
take data quality into consideration.
The authors of that work stress the need
for more quality data rather than quantity data.
To achieve this, they advise researchers to provide a clear description
of the procedures they follow
prior to their analysis and data archiving.
They also encourage the use of automated tools for assessing quality
and the use of sensitivity analysis.

Another related publication is Cheikhi and Abran's~\citeyearpar{CA13}
survey on data repositories.
They noticed that the lack of structured documentation
of \textsc{promise} and \textsc{isbsg} repositories hampered researchers' attempts
to find specific types of data collection.
To address this problem, they supplemented these data collections
with additional information, such as the subject of the study,
the availability of data files and of further descriptions,
and also their usefulness for benchmarking studies.
Information on each study's subject was analyzed
following the corresponding classification of the data studies,
reflecting the classification we subsequently perform on the {\sc msr} data papers
(Section~\ref{sec:rq1}).

In the field of Systems and Software Engineering,
the 13-part study series by Glass et al.~\citeyearpar{Gla94,WTGB11}
assesses scholars and institutions based on the number
of papers they have published in related journals.
The majority of the studies span a five-year period
covering overall the years 1993--2008.
The progress of the study results indicates
that the top three institutions up to 2003 were mainly from the \textsc{usa}
and involved an equal number
of industry research centers and academic institutions.
Since 2004,
the top three institutions are mainly academic
from Korea, Taiwan, and, lastly, Norway.
This change is also observed in the entire list of the 15 top-ranked institutions
presented in the studies.
\textsc{usa} was first in number of top-ranked institutions up to 2002,
followed by Asia-Pacific, Europe, and, lastly, Middle East.
Middle East has disappeared from the list since 2001.
Additionally,
the Asia-Pacific institutions have surpassed \textsc{usa}'s since 2003,
setting Europe in the last place.
Regarding the type of the top-ranked institutions of the list,
during the years 1993--2008,
an average of 82\% were academic institutions,
as opposed to the remaining 18\% which were industry research centers.

A second evaluation of the \textsc{isbsg} software project repository
was carried out by Almakadmeh and Abran~\citeyearpar{AA17}.
Their purpose was to assess the repository from the Six Sigma
measurement perspective and to correlate this assessment
with software defect estimation.
They found that the \textsc{isbsg} Microsoft Excel data extract
contains a high ratio of missing data
within the fields related to the total number of defects.
They consider this outcome a serious challenge,
especially for studies that use the particular data set
for software defect estimation purposes.

The analysis on the Search Based Software Engineering (\textsc{sbse})
publications~\citep{FS11} is the first bibliometric research
of this community, covering a ten-year list of studies,
from 2001--2010.
The evaluation focuses on the categories
of \emph{Publication}, \emph{Sources}, \emph{Authorship}, and \emph{Collaboration}. 
Estimations of various publication metrics are included
for the following years.
Along with the metric forecasting, the authors also studied
the applicability of bibliometric laws in \textsc{sbse},
such as those by Bradford~\citeyearpar{Bra85}
and Lotka~\citeyearpar{Lot26}.

In the same context, Harman et al.~\citeyearpar{HMZ09}
assessed research trends, techniques and their applications in
\textsc{sbse}.
They classified \textsc{sbse} literature
in order to extract specific knowledge
on distinct areas of study.
Then, they performed a trend analysis, which supplied them
with information on activity in these areas.
Finally, for each area of study,
they recognized and presented opportunities for further improvement,
and avenues for supplementary research.

The work of Gu~\citeyearpar{Gu04} is another interesting
bibliometric analysis. The main point of evaluation in this study
is the productivity of authors in the field of Knowledge Management
(\textsc{km}).
To conduct the analysis, Gu collected articles published
in the (former) \textsc{isi} Web of Science\footnote{\url{https://www.webofknowledge.com}} from 1975--2004.
He then recorded all unique productive authors,
along with their contribution and authorship position,
in order to examine their productivity and degree of involvement
in their research publications.
The results indicate that 86\% of the authors
have only written one publication.
As far as citation frequency is concerned,
Gu demonstrates its significant correlation with the reputation
of the journal it has been published in.
On the other hand, his findings reveal no correlation
between \textsc{r\&d} expenditures and research productivity
or citation counts.

In the field of Requirements Engineering,
Zogaan et al.~\citeyearpar{ZSMA17} conducted a systematic literature review
on 73 data sets used for software traceability over a fifteen-year period,
between 2000--2016.
Using both manual and automated methods they selected studies
that have used data sets, case studies, or empirical data,
to develop, validate, train, or test traceability techniques.
Analyzing these studies they identified
that healthcare and aerospace are the two most frequent domains
represented by traceability data sets.
The majority of the data sets are \textsc{oss},
followed by academic and industrial.
Concerning availability,
almost 40\% of the data sets examined are not available for reuse,
originating mainly from industry and academia.
On the contrary, almost all \textsc{oss} data sets are available.
To assess the quality of traceability data sets,
the authors designed a framework consisting of quality characteristics
such as availability, licensing, completeness, trustworthiness,
and interpretability.

\section{Methods}
\label{sec:methods}

We framed our investigation on the usage of \msr\ data papers in terms
of the following research questions.
\begin{description}
\item[RQ1] \emph{What data papers have been published?}
We answer this by finding all data papers published in
the \msr\ proceedings by hand, and further elaborate by manually
clustering them based on the year of publication and the content of the data sets.
\item[RQ2] \emph{How are data papers used?}
We answer this by collecting all citations to
\msr\ data papers by hand and manually classifying them according to their subject
and authors.
\item[RQ3] \emph{What is the impact of published data papers?}
We answer this through the statistical analysis and visualization
of the citations and their slicing according to their type.
\item[RQ4] \emph{What is the community's opinion
regarding data paper publication and use?}
We answer the following subquestions through a web-based survey study
on 108 authors and users of data papers.
\begin{description}
\item[Q4.1] \emph{What motivates people to produce a data paper?}
\item[Q4.2] \emph{How much effort is required to produce a data paper?}
\item[Q4.3] \emph{What are the characteristics of a useful data set?}
\item[Q4.4] \emph{What characteristics prevent a data set from being used?}
\item[Q4.5] \emph{What direction should data papers follow in the future?}
\end{description}
\end{description}
To answer the above research questions,
we performed a mixed methods study.
In order to ensure consistency of the manual processes
that we performed for
the analysis of the research usage of the data papers
(Section~\ref{sec:citation-methods}),
the clustering of them
(Section~\ref{sec:data-paper-methods}),
and the classification of their strong citations
(Section~\ref{sec:citation-methods}),
we employed certain guidelines for
systematic literature reviews
and systematic mapping studies.
Furthermore,
to answer RQ4, we performed a survey study
employing survey research principles
(Section~\ref{sec:survey}).

\subsection{RQ1: Data Paper Collection and Clustering}
\label{sec:data-paper-methods}

We first obtained all data papers
of the proceedings
of the International (Working) Conference
on Mining Software Repositories (\msr).
By the term \emph{data papers} we refer to all papers
included in the data showcase track of the \msr\ proceedings,
as well as other papers from older proceedings
that primarily provide a data set
---
consider e.g. Conklin et al.'s~\citeyearpar{CHC05} collection
of \textsc{floss} data and analyses.

To acquire the aforementioned papers,
we searched through the programs
of the \msr\ conferences
on their respective websites.
Programs that contained an explicit \emph{Data Showcase} section
immediately informed us
about the data papers of the particular year.
In programs that did not include the aforementioned section,
we manually searched
for potential research offering data sets.
From the gathered studies,
those which genuinely offered complete data sets
were included in our data paper archive.

Following the collection, we sorted the data papers
into distinct clusters according to their topic
by combining methods of two prominent studies.
From the systematic mapping studies
in software engineering~\citep{PFMM08},
we applied the classification scheme of abstract keywording.
In addition, we followed two data extraction approaches
suggested in the work of~\citet{BKBT07}
on systematic literature review
within the software engineering domain.
The first approach introduces the use
of two reviewers performing individually the data extraction process
and discussing their disagreements,
while the second approach proposes the use
of a data extractor and a data checker.

According to the above methods,
the first and third authors of this paper
individually labeled all data papers with keywords.
For each data paper,
the two authors read the abstract and extracted keywords
related to the content of the data set
provided by the particular data paper.
For ambiguous abstracts
that hampered the extraction of meaningful keywords,
the two authors also studied the introduction
or conclusion sections of the paper~\citep{PFMM08}.
The keywords were either \emph{in vivo}~\citep{GS67} --- when
representative phrases could be extracted as is
from the aforementioned sections,
or otherwise constructed by the authors.

Following the individual keyword-labeling process,
the two authors met, discussed their keywords
and agreed on a final set of keywords
by refining, merging, and renaming the initial ones.
In this way,
a structured keyword set was formed
consisting of the general keywords
\emph{code} and \emph{people}
(after observing that all individual keywords
could be divided into these two groups),
and more specialized keywords.
\emph{Code} and \emph{people} were used
to signify whether a data paper mostly targeted
the software development process
or the human factor respectively,
while multiple specialized keywords could be assigned to it.
Using the latter keyword set,
the same authors repeated the labeling process
for the data papers together.
Supplementary keywords that appeared
during the second round of keyword assignment
were added to the final keyword set.
Once the second round was completed,
the two authors grouped together the conceptually related keywords.
Through this process,
the clusters of data papers were formed
and then named accordingly.

Finally,
the first and third authors assigned each data paper
to the most conceptually relevant cluster
(i.e., in case a paper could be assigned to more than one cluster,
the authors selected the one they considered most descriptive of its content).
To ensure the correct mapping of the papers to the clusters,
the second author also assessed the cluster assignments,
and then discussed and resolved his objections
with the first and third authors.
The agreement rate of the second author with the first and third authors
was 91\%.
From the 9\% of the disagreements,
57\% were resolved in favor of the second author,
while the remaining 43\% were resolved in favor of the other two authors.

\subsection{RQ2: Data Paper Use Identification and Classification}
\label{sec:citation-methods}

To conduct the analysis on the research usage of the data papers,
we implemented the \emph{Identification of Research}
and \emph{Study Selection} processes,
as proposed in Kitchenham's~\citeyearpar{Kit04} work on procedures
for performing systematic reviews.

The identification of research was made through widely used
and established platforms that provide citation data:
\emph{Google Scholar},\footnote{\url{https://scholar.google.com/}}
\emph{Scopus} --- Elsevier's abstract and citation database\footnote{\url{https://www.scopus.com/}} and the
\emph{ACM Digital Library}.\footnote{\url{https://dl.acm.org/}}
Most research papers that were not publicly available
were provided to us through personal communication with the authors.

After collecting the citations of a particular data paper,
we followed the study selection process.
Specific criteria were applied to the collected research,
in order to ensure quality and validity of our analysis.
First, we applied the whitelisting practice
and kept studies of conference proceedings,
journal articles, master's and doctoral theses,
books, and technical reports.
Studies published in multiple venues,
such as conference publications that are later published in journals
--- e.g. Krueger et al.'s \citeyearpar{KSAB18, KSAB19} study
on the usage of cryptographic \textsc{api}s,
were only listed once.
Priority was given sequentially to books, journal articles,
conference proceedings, reports, and, lastly, theses.
We additionally decided to retain only studies written
in the English language,
due to its widespread adoption for scientific communication.

The main criterion for retaining citing studies was their
actual use of the data sets of the papers they had cited.
We term these \emph{strong} citations.
Research that solely referred to a data paper
without using its data set was not taken into account in our study.
A representative example of a non-strong (\emph{weak}) citation is the study
of repository badges in the npm ecosystem~\citep{TZKV18},
which has cited the collection of social diversity attributes
of programmers~\citep{VSF15a},
although it has not used its data.
Weak citations were manually analyzed to determine
the most common types of uses of data papers in these cases.

The process of citation collection and segregation into strong and weak
was held from July to November 2018.
For data papers that had been strongly cited
by at least one of their authors,
we divided their strong citations into three categories.
The first category contains references
to the data papers made by their first author.
The second category includes strong citations made
by at least one co-author of the respective data paper.
The remaining references that were not made by any
author of the particular data paper were placed in the third category.

Furthermore, we classified the collected strong citations
according to the knowledge areas
of the Guide to the Software Engineering Body of Knowledge~\citep{SWEBOK14}
(\textsc{swebok}).
Again, we followed a combination
of the two data extraction approaches of \citet{BKBT07}
described in Section~\ref{sec:data-paper-methods}.
The first and second authors of this paper individually assigned
each strong citation to a particular \textsc{swebok} knowledge area
after reading their abstract.
Similarly to Section~\ref{sec:data-paper-methods},
in cases of ambiguous abstracts,
they also read the introduction or conclusion sections.
Next, the two authors met, discussed,
and partially resolved their disagreements.
After that,
16\% of the total citation categorizations remained conflicting.
These disagreements were resolved by the paper's last author
who again selected among all knowledge areas.
The selections of the first two authors
were not divulged to him to avoid bias.
Through this process,
30\% of the pending disagreements were resolved
in favor of the first author,
and 15\% in favor of the second author.
Hence, the overall agreement rate between the last and the first two authors was 45\%.
The last author's opinion prevailed over the first two authors'
in the remaining 55\% of disagreements,
where all three assigned knowledge areas were conflicting,
due to his long experience in software engineering
and his extended familiarity with the \textsc{swebok} knowledge areas.

\subsection{RQ3: Citation Analysis}
\label{sec:citation-analysis}

To assess in an objective manner the impact of \msr\ data papers compared to other
\msr\ papers, we collected all \msr\ papers and coupled them with citation data
provided by Scopus.
This process differs from the one described in the preceding
Section~\ref{sec:citation-methods},
because citations are not manually evaluated regarding actual use,
and are retrieved only from a single source (Scopus).
Consequently, the collected metrics are only appropriate for assessing
relative rather than absolute impact.

We first created a data set of all 1267 \msr\ papers by downloading the
complete \textsc{dblp} computer science bibliography database,\footnote{\url{https://dblp.org/}}
and filtering its \textsc{xml} records to obtain only those whose \emph{booktitle}
tag contained \emph{MSR}.
We split the \msr\ papers at hand into two sets:
data papers (as determined in Section~\ref{sec:data-paper-methods})
and the rest.
We also split the \msr\ papers by year to simplify the selection of samples.

Furthermore,
we created a collection mirroring the yearly distribution of data papers
in order to compare in a fair manner citations to data papers
against citations to other \msr\ papers.
We created this collection as follows.
For each year in which $N$ data papers were published, we randomly chose
$N$ non-data papers from the \msr\ papers published in the same year.

To assess research building on data papers, we also created a set of
\msr\ papers that cite \msr\ data papers.
We did this by calculating the intersection between all \msr\ papers
and the papers that use them
(as determined in Section~\ref{sec:citation-methods}).
Although this new set of papers citing data papers is not
exhaustive (it only contains \msr\ papers), it allows us
to compare the citation metrics of these papers against those
of a known tractable population, namely \msr\ papers as a whole.

We then used the Scopus \textsc{rest api} to obtain the number of times
each \msr\ paper was cited.
The citation data obtained in this step are not comparable
with those we obtained through the widespread search and manual
filtering described in Section~\ref{sec:citation-methods},
because they may be associated with false positives and false negatives.
However, they allow comparisons to be made between different \msr\ sets,
because all citation metrics are obtained through the same
methods employed by Scopus and all probably suffer
from the same types of bias.

Finally, we joined the Scopus citation data with the sets obtained
in the previous steps.
We then calculated simple descriptive statistics for the citation counts
of the following sets:
\begin{itemize}
\item all \msr\ data papers,
\item a sample of \msr\ non-data papers mirroring the yearly distribution of
data papers,
\item all \msr\ non-data papers for years in which data papers were published,
and
\item \msr\ papers citing \msr\ data papers.
\end{itemize}

\subsection{RQ4: Survey Planning, Execution, and Analysis}
\label{sec:survey}

To conduct our survey study on authors and users of data papers
in order to explore our community's view
regarding data paper publication and use,
we followed the set of ten activities
introduced in Kitchenham and Pfleeger's~\citeyearpar{PK01,KP02a,KP02b,KP02c,KP02d,KP03}
six-part series of principles of survey research.

\textbf{Survey design.}
We adopted a cross sectional,
case control observational study design
(i.e., participants were surveyed about their past experiences
at a particular fixed point in time),
which is typical of surveys in software engineering~\citep{KP02a}.
The goal of this study was
\emph{to obtain further insights on the production,
use, and future desired direction of data papers}.
Hence, we framed the objectives of our survey in terms of the following questions.
\begin{description}
\item[Q1] \emph{What motivates people to produce a data paper?}
\item[Q2] \emph{How much effort is required to produce a data paper?}
\item[Q3] \emph{What are the characteristics of a useful data set?}
\item[Q4] \emph{What characteristics prevent a data set from being used?}
\item[Q5] \emph{What direction should data papers follow in the future?}
\end{description}
To prevent a low response rate due to the summer vacation period
that co-occurred with the study preparation,
the survey was scheduled to run in early September 2019.

\textbf{Survey sample.}
The survey was conducted on two different samples
in two different time periods --- September 2019
and January 2020.
The reason behind the second conduction was to include more strong citation authors.
In the first conduction,
all primary authors of the 81 data papers comprised our sample,
along with an equal set of primary authors of strong citations.
The unique primary authors of data papers were in total 71.
For the selection of the 71 (out of 419) primary authors
of strong citations,
we implemented the probabilistic sampling method
of simple random sample~\citep{KP02d}.
For that purpose,
we used the random number generator
of the Python 3.7 \emph{random} library
with the default seed value and replacements.
From the collected candidate respondents,
14 were primary authors of both data papers and strong citations,
leading to a total of 128 unique candidates
(instead of 142).
In the second conduction,
the remaining primary authors of strong citations comprised our sample
excluding the ones already included in the first sample.
In this manner,
our second sample was composed of 189 unique candidates,
while the overall sample size was 317.

\textbf{Survey instrument.}
The survey questionnaire
was organized into eight sections.
In the first section participants specified
whether they were authors of data papers,
along with the number of their data paper publications in \msr\
and in any other venue.
The next five sections were organized
according to the objective questions,
and were composed of both mandatory and optional
open-ended, multiple choice,
and Likert scale questions.
We intentionally used even Likert scales to
force participants to make a choice~\citep{GSB16}.
To address Q1,
data paper authors applied on a four-level Likert scale
the extent to which a set of specific attributes motivate them
to produce a data paper.
There was also an open-ended question
for additional motivational factors.
For Q2,
data paper authors specified through a multiple choice question
the effort-months they need on average
to produce a data paper.
For Q3 and Q4,
both authors and users of data papers evaluated on a four-level Likert scale
the importance of a set of particular characteristics
for the selection or avoidance of a data set for their research.
The same characteristics were included in both questions.
In our view, an attribute considered as useful to some extent
is not necessarily considered as discouraging to the exact same extent,
thus affecting the overall ranking of the characteristics.
An open-ended question was included
for additional useful or preventing attributes of data sets.
To address Q5,
respondents listed through open-ended questions
what data papers they would like to see published in the future,
and from what sources new data could be derived.
The following section included demographic questions
aiming to assess the diversity of the responses.
Through the final section of the survey
we retrieved feedback regarding the completeness of the questionnaire;
respondents assessed whether the objective questions 
were sufficiently or weakly addressed,
and left their comments.
Finally, they could leave their e-mail address,
to receive a report with the survey results.
To ensure anonymity,
the collected e-mail addresses are not publicly distributed
within the online available data set of survey responses.

\textbf{Survey evaluation.}
To evaluate and refine the questions of the survey,
and to calculate the average time required to complete it,
we initially performed a pilot study.
The sample of the pilot study was composed
of 23 members of our laboratory
(two faculty members, three senior, six associate,
and twelve junior researchers),
four external faculty members, and one more senior researcher.
In total 28 potential respondents constituted the sample of the pilot study.
The pilot study ran from July 25th to August 1st, 2019,
and nine responses were received (32\% response rate).
Respondents of the pilot study were asked
at the beginning of the questionnaire
to complete their current local time.
Through subtraction from the reported completion timestamp
(which was automatically recorded unlike the starting timestamp),
the average time required was calculated at 18 minutes.

\textbf{Survey operation.}
Both the pilot and the final survey were distributed
as a \emph{Google form},
which the candidate participants were invited to complete
through an invitational mail.
Although the mailing process was automated,
it was personalized by addressing the candidates by name,
and by including details on how they were selected
and which of their research papers
(data papers and/or strong citation papers) the survey involved.
The candidates were informed on the average time required
for the questionnaire completion (rounded up to 20 minutes),
along with the goal and the objectives of the survey study. \\
From the total of 317 mails that were sent as part of the final survey,
39 failed to be delivered.
These failures involved twelve e-mail addresses no longer in use,
26 rejected recipients, and one wrong e-mail address.
We consider our final sample a total of 278 potential participants. \\
The final survey ran from September 2nd to 24th, 2019,
and from January 24th to February 16th, 2020.
In both rounds we aimed for a three week duration,
but we would briefly reopen the survey
when candidate participants requested it. \\
A reminder mail was distributed to potential respondents
ten days after the invitational mail in both rounds.
Verified respondents who had either answered our initial mail
or had left their e-mail address in their survey response were excluded
from the recipient list of the reminder mail.
The final survey received 108 responses (39\% response rate
calculated on the basis of the final sample size --- 278).

\textbf{Survey analysis.}
Q3--Q5 were analyzed from the perspective
of authors and users combined,
and of users independently.
We applied manual pair coding~\citep{SPP08}
to summarize the results of the six open-ended questions.
For the first survey conduction,
the first and second authors of this paper applied together codes
to all open-ended responses
following a mixed approach
of line-by-line and sentence-by-sentence coding~\citep{Cha16}.
Next, they combined conceptually-related codes
by generalizing or specializing them,
and integrated them into distinct groups.
For the second survey conduction,
the same authors used the first group of codes
to annotate the new responses~\citep{GSB16}.
In case a response was not connected to any group,
or was connected but further ideas were also introduced,
the authors would apply new codes to it.
At the end,
the generalization-specialization and grouping process was repeated
for the new codes.

\subsection{RQ4: Survey Participants}
\label{sec:participants}

The questionnaire was completed by 108 respondents of various age groups.
Concerning their current occupation,
37\% (24) were academic staff,
24\% (15) worked in industry,
19\% (11) were post doctoral researchers, and
19\% (5) were doctoral students.

From the 108 respondents of the survey,
30\% (32) were primary authors only of strong citation papers,
as opposed to the remaining 70\% (76) who were primary authors of data papers,
with a portion of them also being authors of strong citation papers.
From the 76 data paper authors,
one's data papers have not been published in any venue, 
42\% (32) have published only one data paper,
26\% (20) have published two,
11\% (8) of the authors have published three,
5\% (4) have published four,
and three authors own five, six, and seven publications respectively.
Furthermore, 3\% (2) of the authors have published eight data papers,
followed by another 3\% (2) with ten publications.
Lastly,
four authors own eleven, 20, 25, and 30 data paper publications respectively.
Concerning data paper publications in the \msr\ conference,
22\% (17) of the authors have no \msr\ data papers,
55\% (42) have one,
15\% (11) of the authors have two \msr\ publications,
3\% (2) own three publications in \msr,
4\% (3) own four,
and one is the primary author of five \msr\ data papers.

\section{Results}
\label{sec:results}

The findings of our study are framed in respect
to the four research questions.

\subsection{RQ1: What data papers have been published?}
\label{sec:rq1}

\begin{table}[t]
\renewcommand{\arraystretch}{1.4}
\caption{\label{tab:data-papers-by-year}MSR Data Papers by Year\newline
There has been a significant rise in the number of data papers
since 2013, the year that the MSR data showcase track was established.}
\centering
\begin{tabular}{lL{10.5cm}}
\hline
Year & Data Papers \\
\hline
\input{data-papers-by-year}
\hline
\end{tabular}
\end{table}

\begin{figure}[t]
\centering
\includegraphics[width=0.7\textwidth, keepaspectratio]{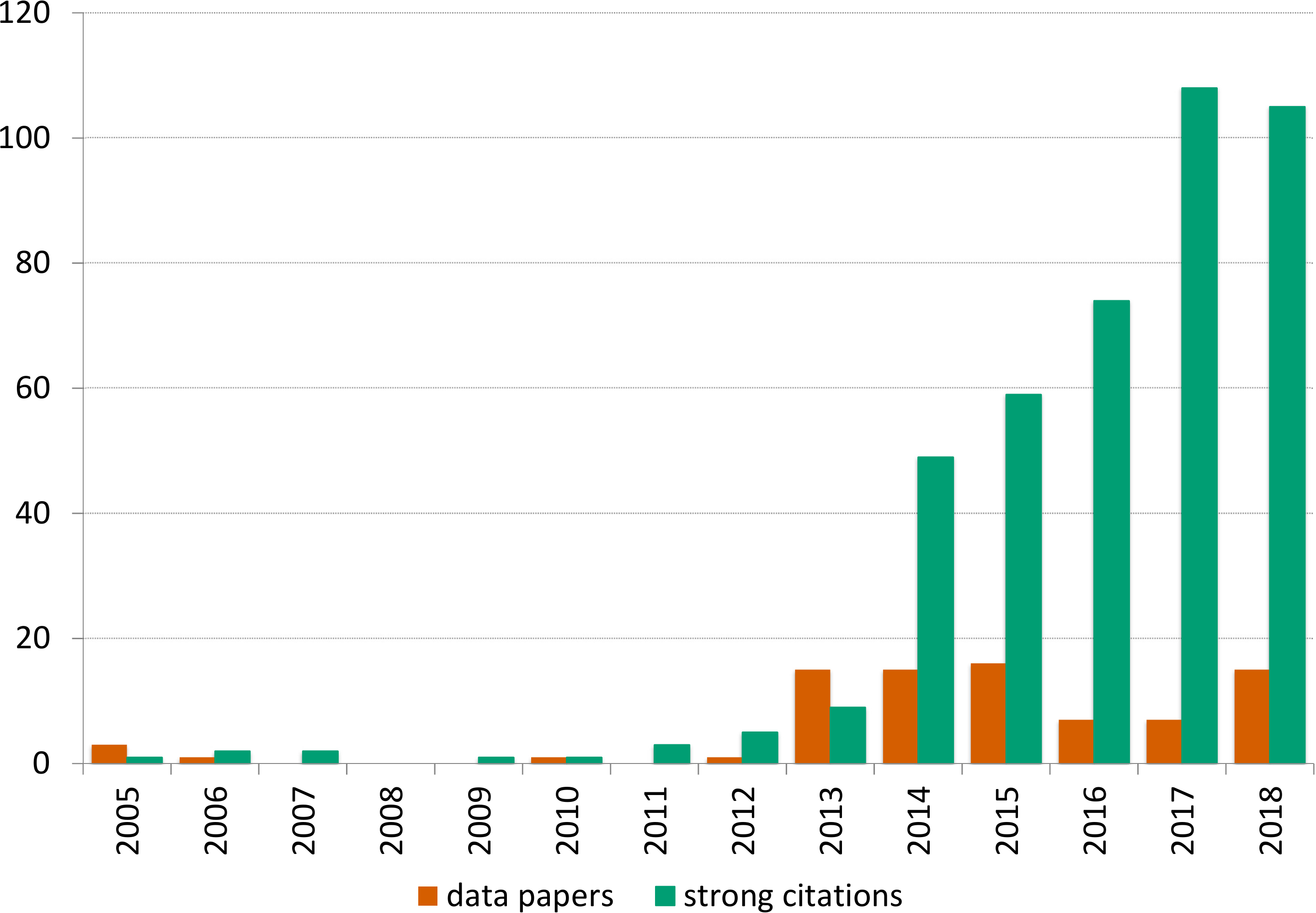}
\caption{Timeline of the data papers and the strong citations.
Each year depicts the number of data papers
published in the particular year and the number of studies
published in the particular year that are based on any data paper. The
number of strong citations to data papers is constantly rising,
indicating that the concept of data papers has long-term value.}
\label{fig:yearly-papers}
\end{figure}

\begin{table*}[t]
\renewcommand{\arraystretch}{1.3}
\caption{\label{tab:data-paper-clusters}Data Paper Clusters and Strong Citations\newline
Publications providing VCS primary and derived data are the most frequent
data papers and the most often strongly cited ones.}
\centering
\resizebox{\textwidth}{!}{
\begin{tabular}{l r r r r r}
\hline
Cluster & Data Papers & Str. cited DPs & Non-cited DPs & Str. Citation Ratio (\%) & Str. Citations \\
\hline
VCS Primary and Derived Data & 29 & 20 & 9 & 69 & 312 \\
Software Faults, Failures, Smells & 17 & 11 & 6 & 65 & 42 \\
Software Evolution & 11 & 6 & 5 & 55 & 17 \\
Group Dynamics & 9 & 7 & 2 & 78 & 41 \\
Computational Linguistics & 7 & 5 & 2 & 71 & 20 \\
Software Models & 3 & 2 & 1 & 67 & 5 \\
Computing Education and Programming Practices & 3 & 1 & 2 & 33 & 1 \\
Enhanced Developer Data & 2 & 1 & 1 & 50 & 2 \\
\hline
\textbf{Total} & 81 & 53 & 28 & 65 & 440 \\
\hline
\end{tabular}
}
\end{table*}

We identified the 81 data papers presented in Table~\ref{tab:data-papers-by-year}.
These comprise about 15\% of the 507 papers
published in the \msr\ conference
in the years when data papers appeared.
The timeline of the data papers and the research based on them
is depicted in Figure~\ref{fig:yearly-papers}.
For each year, the number of published data papers is showcased,
along with the number of studies published in the particular year,
which have been based on any of these data papers.
(It should be noted that this is not a cumulative graph;
each year's outcome is independent of the previous.)
There has been a significant rise
in the number of data papers since 2013,
which is the year
when the data showcase track was founded~\citep{ZMSD13}.
Until then,
2005 was the year with the most data papers.
The smallest number of data showcase research papers --- seven --- was published
in 2016 and 2017.
Nevertheless, 2018 indicates a double increase
in data publications --- 15 (see Table~\ref{tab:data-papers-by-year}).

From the clustering of the data papers,
as described in Section~\ref{sec:data-paper-methods},
eight data clusters emerged.
Table~\ref{tab:data-paper-clusters} shows for each cluster
the number of data papers it comprises,
the number of strongly cited and non-cited data papers,
and the strong citations that have been made to them.
We consider as \textit{non-cited}
the data papers with either weak citations
or no citations at all.
The clusters are sorted
in descending order according
to the number of data papers
they contain.

\emph{VCS Primary and Derived Data} preponderate.
The particular cluster consists of 29 studies
that provide Version Control System (\textsc{vcs})
raw or processed data,
along with descriptive statistics and analyses.
The collection of Java source code of the Merobase
Component Finder project~\citep{JHSA13} is part of this cluster.

\emph{Software Faults, Failures, Smells} concern 17 data sets
of security failures, software inconsistencies,
and bad programming practices
detected in a variety of software applications and ecosystems.
For instance, Vulin\textsc{oss} offers a data set
of security vulnerabilities in open-source systems~\citep{GMS18}. 

\emph{Software Evolution}
involves eleven collections with information
on the evolution of artifacts, such as operating systems~\citep{Spi18},
software architectures~\citep{WY15}, and software packages~\citep{BAKO14a}.

The cluster of \emph{Group Dynamics} is composed of nine data papers
that focus on social networks~\citep{MK13},
code reviewing~\citep{MBR13}, project roles~\citep{Squ13b},
and social diversity in programming teams~\citep{VSF15a}.

Seven data papers were grouped together
due to their common theme of facilitating
studies related to natural language processing and information extraction~\citep{PML15, BLPH13}.
These papers constitute the \emph{Computational Linguistics} cluster.

\emph{Software Models} provide simplified visual representations
of software processes, such as simplified syntax trees~\citep{PANM16a}
and \textsc{uml} models~\citep{RHHC17}.

Papers that share records regarding novices' and experts'
programming practices and abilities
--- e.g. the list of Scratch programs of students~\citep{AHMR17} ---
were classified in \emph{Computing Education and Programming Practices}.
The aim of this cluster is to facilitate studies on computing education.

The last defined cluster contains papers
offering \emph{Enhanced Developer Data},
such as screen and real names extracted from Twitter~\citep{Squ13a}
and personal characteristics (e.g. gender, age, civil status, nationality), education and level of English, and professional status~\citep{RASV14}.
Only two papers represent this cluster,
however the uniqueness of their data sets segregates them
from the other clusters.

\subsection{RQ2: How are data papers used?}
\label{sec:rq2}

\begin{table*}[t]
\renewcommand{\arraystretch}{1.3}
\caption{\label{tab:top-five-data-papers}Top Five Data Papers
in Number of Strong Citations\newline
The most strongly cited data paper offers a collection
of primary and derived data extracted from GitHub.}
\centering
\resizebox{\textwidth}{!}{
\begin{tabular}{m{6cm} l r l r}
\hline
Title & Data Paper & Year & Cluster & Str. Citations \\
\hline
The GHTorrent Dataset and Tool Suite & \citep{Gou13} & 2013 &
VCS Primary and Derived Data & 165 \\
AndroZoo: Collecting Millions of Android Apps for the Research Community & \citep{ABKL16} & 2016 & VCS Primary and Derived Data & 57 \\
Lean GHTorrent: GitHub Data on Demand & \citep{GVSZ14a} & 2014 &
VCS Primary and Derived Data & 24 \\
Who Does What During a Code Review? Datasets of OSS Peer Review Repositories & \citep{HKYC13} & 2013 & Group Dynamics & 16 \\
The Maven Repository Dataset of Metrics, Changes, and Dependencies & \citep{RVV13} & 2013 & VCS Primary and Derived Data & 12 \\
The Eclipse and Mozilla Defect Tracking Dataset: A Genuine Dataset for Mining Bug Information & \citep{LPD13} & 2013 & Software Faults, Failures, Smells &
12 \\
The Emotional Side of Software Developers in JIRA & \citep{OMDT16} & 2016 &
Computational Linguistics & 12 \\
\hline
\end{tabular}
}
\end{table*}

The 81 \msr\ data papers are associated with 1169 citations to them,
coming from 982 distinct studies
(some studies cite multiple data papers).
Out of the 1169 citations, 440 (419 distinct studies) use the data sets provided by the data papers (\emph{strong} citations).
The remaining 729 citations (610 distinct studies) refer to data papers
without utilizing the particular data sets (\emph{weak} citations).
We were able to obtain most citations
from digital libraries and the web.
Six citations that were publicly unavailable were received from their respective authors
through personal communication,
as stated in Section~\ref{sec:citation-methods},
but no access was obtained for another three.
(These three studies have been excluded from the total citations.)
Table~\ref{tab:top-five-data-papers} depicts the most strongly cited data papers.

Through manual analysis
we found that the most common uses of weak citations were
mentioning the work as an example ($n=524$, 77\%),
attributing a work's statement ($n=59$, 9\%),
using the work's methods ($n=47$, 7\%),
presenting the study as related work ($n=12$, 2\%), and
reporting obtained statistics ($n=13$, 2\%).

\renewcommand{\sidecaptionsep}{0.6cm}
\begin{SCtable}[][b]
\renewcommand{\arraystretch}{1.2}
\caption{\label{tab:strong-citation-categories}Areas of Strong Citing Studies\newline
The studies that strongly cite data papers span the SWEBOK knowledge
areas fairly unequally.}
\resizebox{0.68\columnwidth}{!}{
\begin{tabular}{l r r}
\hline
SWEBOK Knowledge Area & Studies & Percentage\\
\hline
Software Maintenance & 89 & 21.2 \\
Software Engineering Management & 63 & 15.0 \\
Software Engineering Professional Practice & 57 & 13.6 \\
Software Quality & 55 & 13.1 \\
Software Configuration Management & 46 & 11.0 \\
Software Construction & 43 & 10.3 \\
Software Design & 20 & 4.8 \\
Software Engineering Process & 19 & 4.5 \\
Software Testing & 15 & 3.6 \\
Software Engineering Economics & 6 & 1.4 \\
Software Engineering Models and Methods & 5 & 1.2 \\
Software Requirements & 1 & 0.2 \\
\hline
\end{tabular}
}
\end{SCtable}

Table~\ref{tab:strong-citation-categories} depicts
the classification of the studies based on data papers
according to the knowledge areas of the \textsc{swebok}.
This suggests that research on \emph{Software Maintenance},
\emph{Software Engineering Management},
and \emph{Software Engineering Professional Practice} uses data papers
to a considerable extent.
On the other hand, only a slight portion of research
on \emph{Software Requirements},
\emph{Software Engineering Models and Methods},
and \emph{Software Engineering Economics} is facilitated
by data showcase papers.

\begin{figure*}[t]
\centering
\includegraphics[width=\textwidth, keepaspectratio]{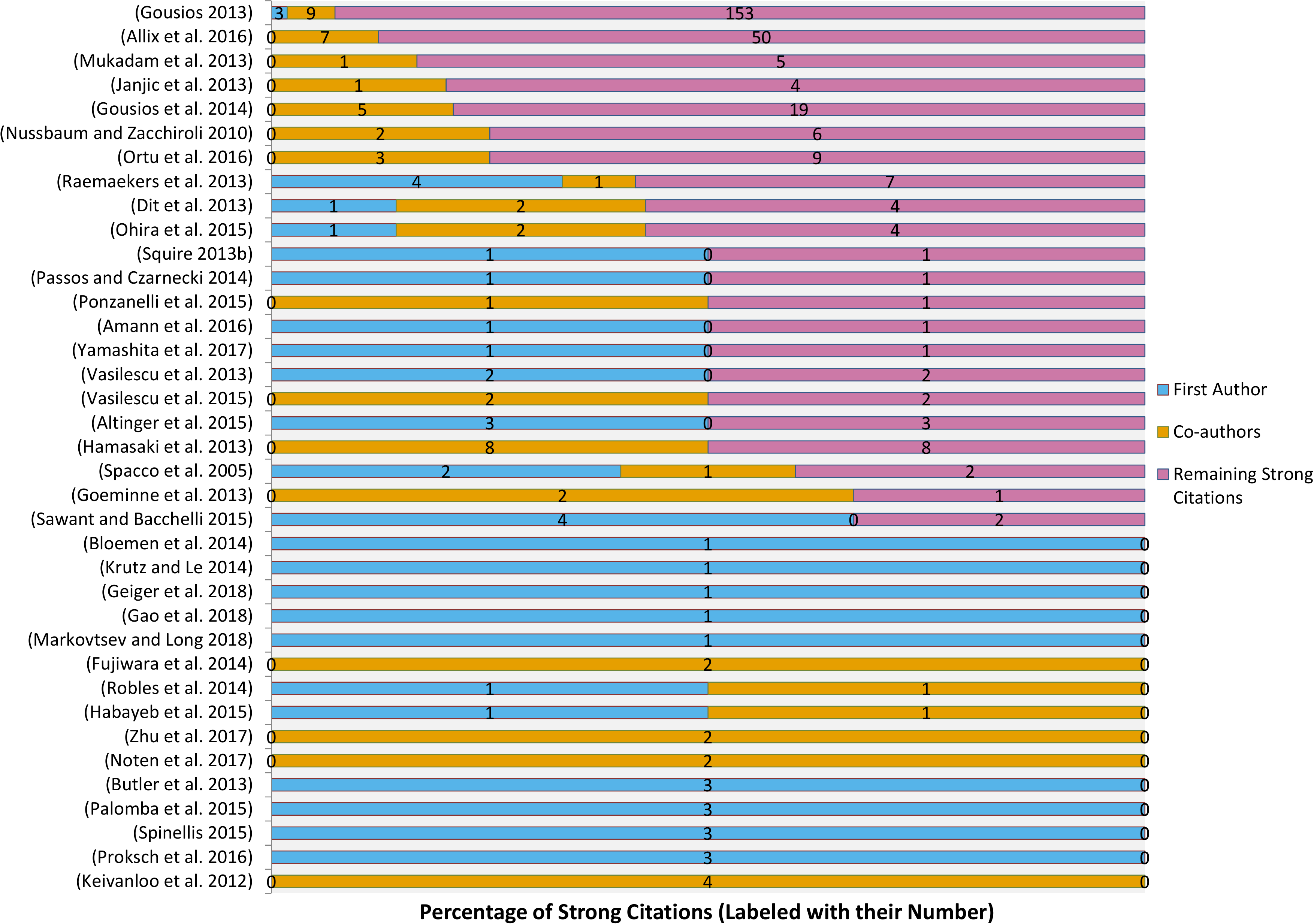}
\caption{Use of data papers by their authors (\%).
Data papers used at least once by the same first author or any of his/her co-authors
are represented by the number of strong citations made by the first author,
the co-authors, and other unrelated teams. From the total 81 data papers,
37 have been used by the teams that authored them.}
\label{fig:self-citations}
\end{figure*}

Furthermore, concerning the use of data papers by their respective authors, 
our findings show that 37 out of the 81 papers have been used
by the teams that authored them.
Specifically, 15 studies have been solely used
either by their first author or his/her co-authors.
Figure~\ref{fig:self-citations} depicts for each data paper
strongly cited at least once by the first author or the co-authors,
the percentage of the uses that stem from the first author,
the co-authors, and other unrelated teams.
The data papers are sorted in ascending order downwards
based on the percentage of the sum of the strong citations
made by the first author and the co-authors.
For instance, 67\% of the strong citations to the collection
of \textsc{api}s usage information~\citep{SB15} were made by the first author.

\subsection{RQ3: What is the impact of published data papers?}
\label{sec:rq3}

The relative impact of published data papers
can be deduced from Table~\ref{tab:citation-metrics},
which compares citations to data and non-data papers,
collected in the way described in Section~\ref{sec:citation-analysis}.
(The three data papers missing from the table are those published in
\msr\ '05, which are not tracked by Scopus.)
The table shows
that data papers are typically cited less often,
compared to others of the \msr\ conference in terms of
the median and average number of references.
This occurs both in terms of yearly-weighted samples and
as a whole.
Also, \msr\ papers that cite data papers appear to be
cited about the same ($\mu=10$, $\bar{x}=15,$ 7)
as other \msr\ papers ($\mu=8$, $\bar{x}=17,$ 0),
meaning that citing an \msr\ data paper does not promise greater popularity concerning incoming citations.

\renewcommand{\sidecaptionsep}{0.6cm}
\begin{SCtable}[][t]
\renewcommand{\arraystretch}{1.1}
\caption{\label{tab:citation-metrics}Citation Metrics by Paper Type\newline
Data papers are typically cited less often, compared to other papers of
the MSR conference, in terms of the median and average number of
references.}
\resizebox{0.6\columnwidth}{!}{
\begin{tabular}{l rrrr}
\hline
Metric	& Data Papers	& Non-DP	& Non-DP	& Citing DP \\
	& 		& (Sample)	& (All)		& \\
\hline
\input{citation-metrics}
\hline
\end{tabular}
}
\end{SCtable}

\renewcommand{\sidecaptionsep}{2.2cm}
\begin{SCtable}[][t]
\renewcommand{\arraystretch}{1.2}
\caption{\label{tab:venues}Venues With Research Based on Data Papers
\newline The majority are top-notch venues, indicating the high quality
of studies that can be performed through data papers.}
\resizebox{0.4\columnwidth}{!}{
\begin{tabular}{l r r}
\hline
Venue & Papers & Percentage \\
\hline
\input{venues}
\hline
\end{tabular}
}
\end{SCtable}

Table~\ref{tab:venues} shows the venues where research that is based on
data papers has been published.
We see that more than a third of the corresponding papers are published
in top-tier conferences and journals.
This showcases the high quality of research that is conducted based
on data papers.
We examined by hand the papers published
in the Computing Research Repository (CoRR),\footnote{\url{https://arxiv.org/corr}}
and found that almost all of them (19) are fairly recent
(published in 2017 or 2018).
This indicates that they are probably archival submissions of material that will
eventually also end up in a conference or journal.

The timeline of the data paper uses is depicted
in Figure~\ref{fig:yearly-papers}.
The strong citations of all data papers were summed up
and illustrated as yearly records.
We see that strong citations have risen since 2014,
which was expected after the data showcase track's introduction in 2013.
Only six studies were identified before the category's establishment.

\begin{figure}[t]
\centering
\includegraphics[width=0.6\textwidth, keepaspectratio]{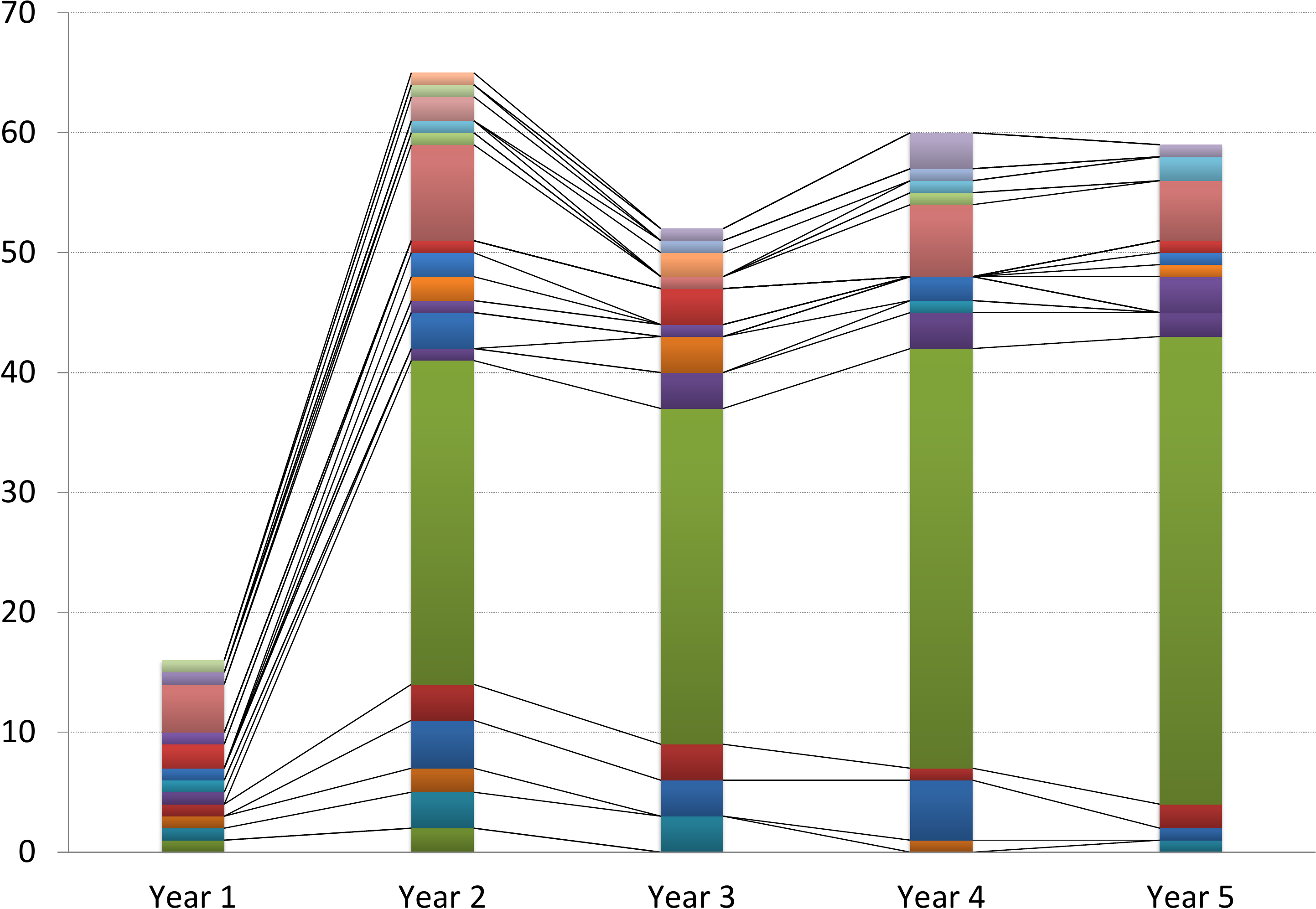}
\caption{Timeline of strong citations to data papers published from 2005--2014
over a five-year window. Each strongly cited data paper is represented
with the same color along the years. The height of each color bar is
relative to the number of strong citations.
Data papers with the most strong citations during the second year of their existence seem to retain their citation number
in the following years, or to obtain even more strong citations.}
\label{fig:five-year-citations}
\end{figure}

In addition, we studied the growth of data paper use
in a five-year window after the data papers' publication,
and imprinted it on Figure~\ref{fig:five-year-citations}.
The limit five was chosen
because it provided us with sufficient insights,
without excluding too many papers
that were less than five years old.
Consequently, we included data studies published in the years 2005--2014.
The majority of them reveal a peak in the number of strong citations
during the second year of their existence,
but appear to have a significant decrease of uses in the following year.
Research based on data papers seems to plateau
after the third year of their life.

\subsection{RQ4: What is the community's opinion
regarding data paper publication and use?}
\label{sec:rq4}

In this section we present
the answers of the survey respondents to the questionnaire
in respect to the objective questions,
and their feedback on the survey study.

\textbf{Q4.1: What motivates people to produce a data paper?}

\begin{figure}[t]
\centering
\includegraphics[width=\linewidth]{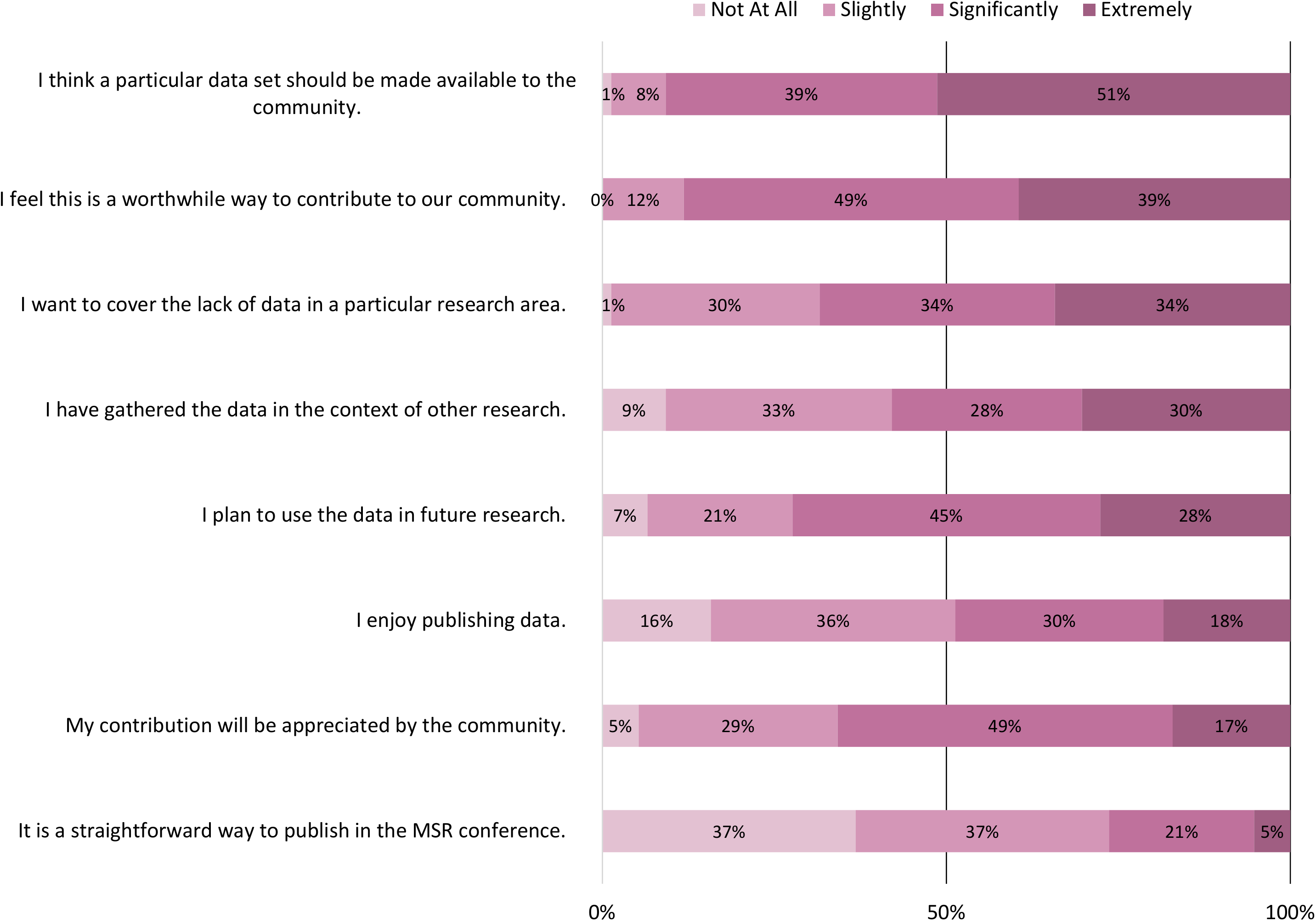}
\caption{Motivational impact of specific attributes for data paper production.
According to the responses of the primary data paper authors,
the motivational impact of a set of eight predefined characteristics
for data paper production is depicted on a four-level Likert scale
and on a percentile basis of the total data paper authors.
The majority of the participants think that particular data sets should be made available,
and feel that publishing data papers is
a worthwhile way to contribute to our community.}
\label{fig:motivational-factors}
\end{figure}

All 76 data paper authors were requested to assess
on a four-level Likert scale the motivational impact
of a set of eight predefined attributes for data paper production.
The collected answers to this question are illustrated
in Figure~\ref{fig:motivational-factors},
where the attributes are sorted downwards in descending order
of the \emph{Extremely} level.
The majority of participants claim to a significant or extreme degree
that publishing data papers is
\emph{a worthwhile way to contribute to our community},
and that their \emph{contribution will be appreciated by the former}.
In addition,
they emphasize significantly that
\emph{certain data sets should be made available to the community},
while others \emph{want to cover the lack of data in a particular
research area}.
To a similar extent of high significance,
the responding authors of data papers
mention that they usually
\emph{have gathered the data in the context of other research},
or \emph{plan to use their published data in future research}.

Apart from the predefined set of attributes,
some participants also stated in the related open-ended question
regarding additional motivational aspects,
the expectancy of obtaining a high number of citations,
the challenging process of data paper production,
validity, transparency, research reusability, and quality improvement
as another set of aspiring characteristics.
Others recognized that data papers provide a more thorough
data set representation and description,
promote open science and potential partnerships,
and reveal new research trends.
A few respondents identified their skillfulness
in data paper production as another valuable factor.

\textbf{Q4.2: How much effort is required to produce a data paper?}

From the 76 data paper authors who responded to the survey,
8\% (6) stated that they need less than an effort-month,
36\% (27) need from one to three effort-months,
34\% (26) require a total of four to six effort-months,
15\% (11) of the authors need from seven to nine effort-months,
7\% (5) need from ten to twelve effort-months,
and one needs more than a year of effort to produce a data paper.

\textbf{Q4.3: What are the characteristics of a useful data set?}

\begin{figure}[h!]
\centering
\includegraphics[width=\linewidth, height=0.45\textheight]{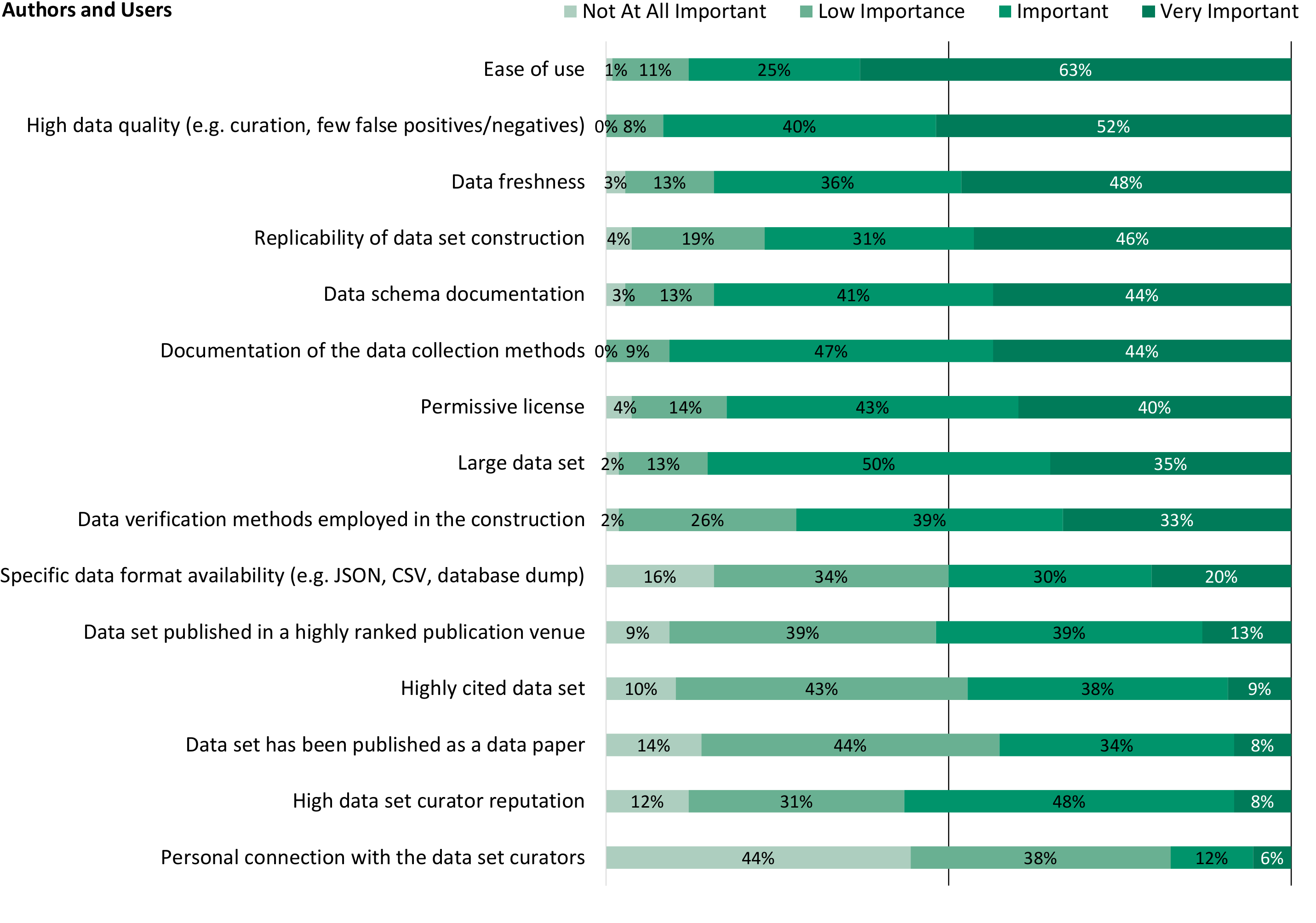}
\includegraphics[width=\linewidth, height=0.45\textheight]{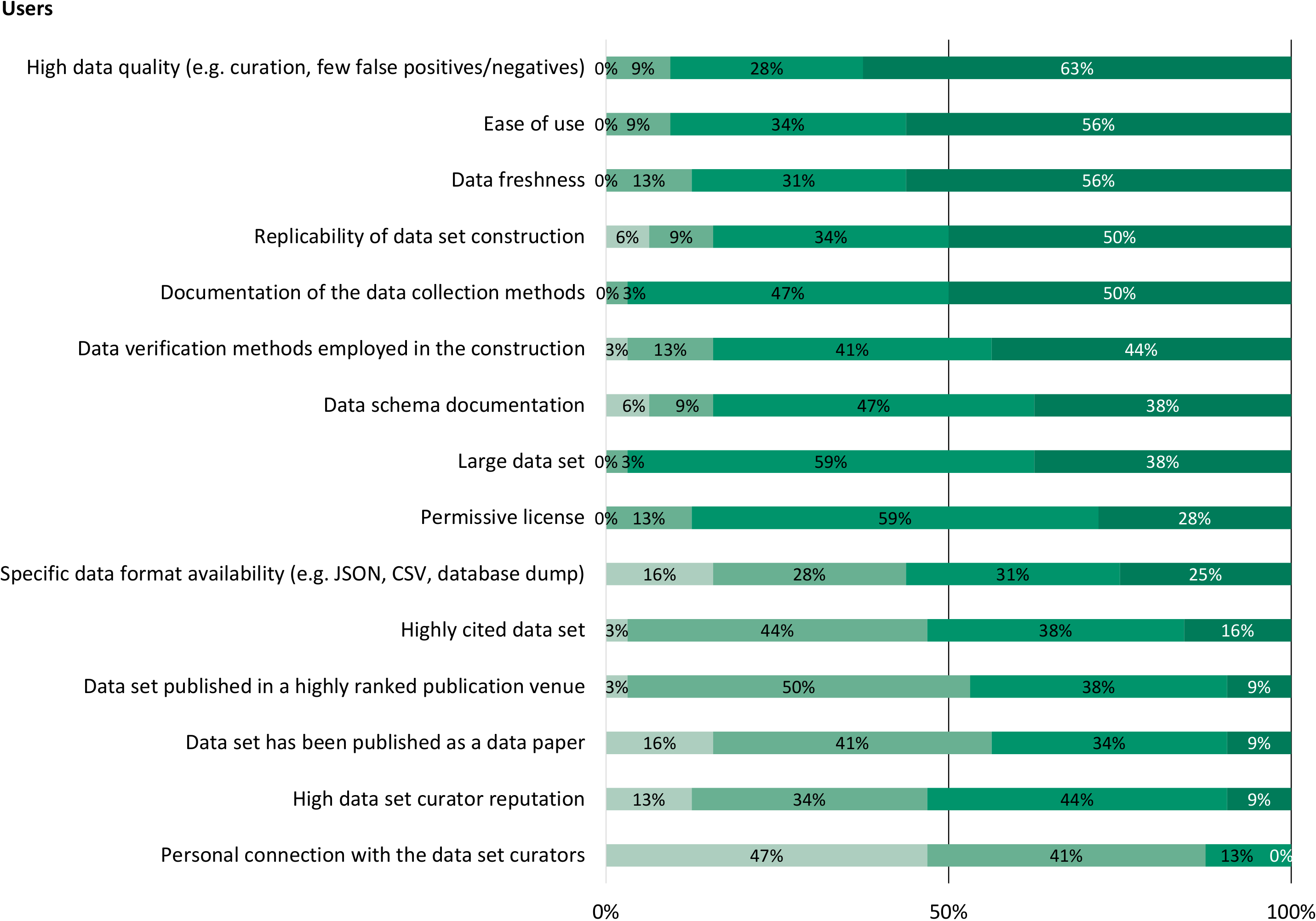}
\caption{Encouraging characteristics in data set selection
for data paper authors and users (top), and users isolated (bottom).
Respondents assessed a set of 15 predefined characteristics
on a four-level Likert scale.
Ease of use and high data quality appear to be the most
important characteristic in data set selection.}
\label{fig:useful-characteristics}
\end{figure}

In Figure~\ref{fig:useful-characteristics},
the assessment of a set of 15 characteristics regarding their importance
in data set selection for research purposes is illustrated.
The characteristics are sorted downwards in descending order
of the \emph{Very Important} level.
All 108 respondents evaluated the particular attributes
on a four-level Likert scale.
The most important characteristics were found to be
\emph{ease of use}
(i.e., the ability of users to effortlessly obtain the information they are after),
\emph{high data quality},
\emph{data freshness},
\emph{replicability of data set construction},
\emph{data schema documentation},
and \emph{documentation of the data collection methods}.
Less important characteristics included
\emph{personal connection with the data set curators},
\emph{data set having been published as a data paper},
and \emph{data set having been highly cited}.
Separating data paper users from authors
in Figure~\ref{fig:useful-characteristics},
we observe a similar ranking of the characteristics.
The main difference is in the first place,
which for users is \emph{high data quality},
as opposed to \emph{ease of use} selected by authors and users combined.

In addition to the above characteristics,
through the complementary open-ended question,
respondents stressed the importance of
existing application examples,
compatibility and extensibility with other data sets,
data balance and integrity,
data completeness, reproducibility, updatability,
and ease of access and filtering.
Participants also valued
modification traceability,
data novelty and diversity,
data validation,
continuous maintenance of data
and support by the curators,
and enhancement of existing data sets.
Furthermore,
documented threats and flaws,
curation of duplicate data (e.g. repository forks) and code clones,
inclusion of timestamps,
goal-oriented data,
anonymity of subjects,
contextuality (i.e., the data set captures its context),
duality (i.e., the data set contains both positive and negative samples where applicable),
and usage metadata of the data set are also appreciated.
For software engineering fields with ongoing radical changes,
such as malware detection,
some remarked that the shelf life
of a data set should be short.

\textbf{Q4.4: What characteristics prevent a data set from being used?}

\begin{figure}[h!]
\centering
\includegraphics[width=\linewidth, height=0.45\textheight]{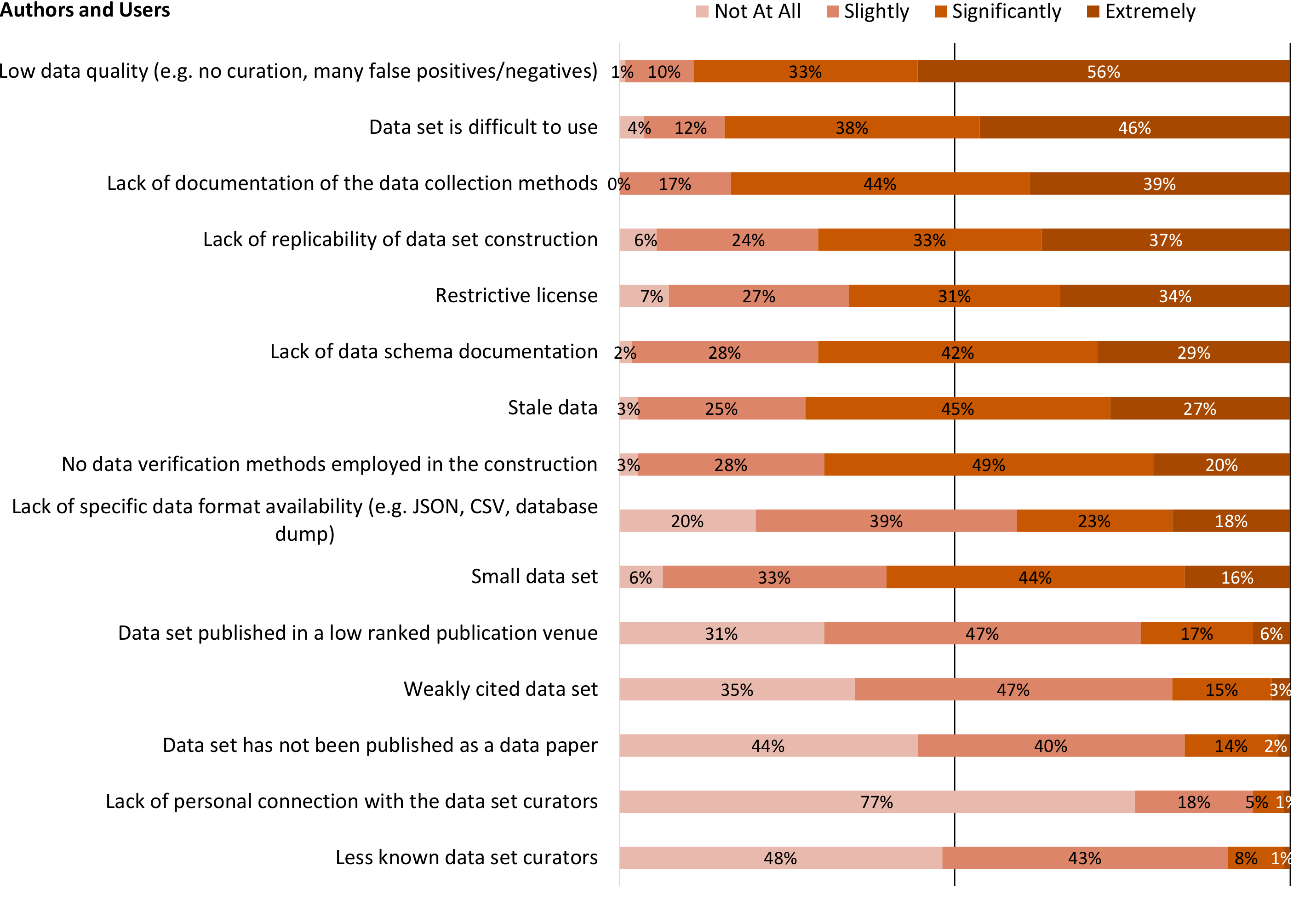}
\includegraphics[width=\linewidth, height=0.45\textheight]{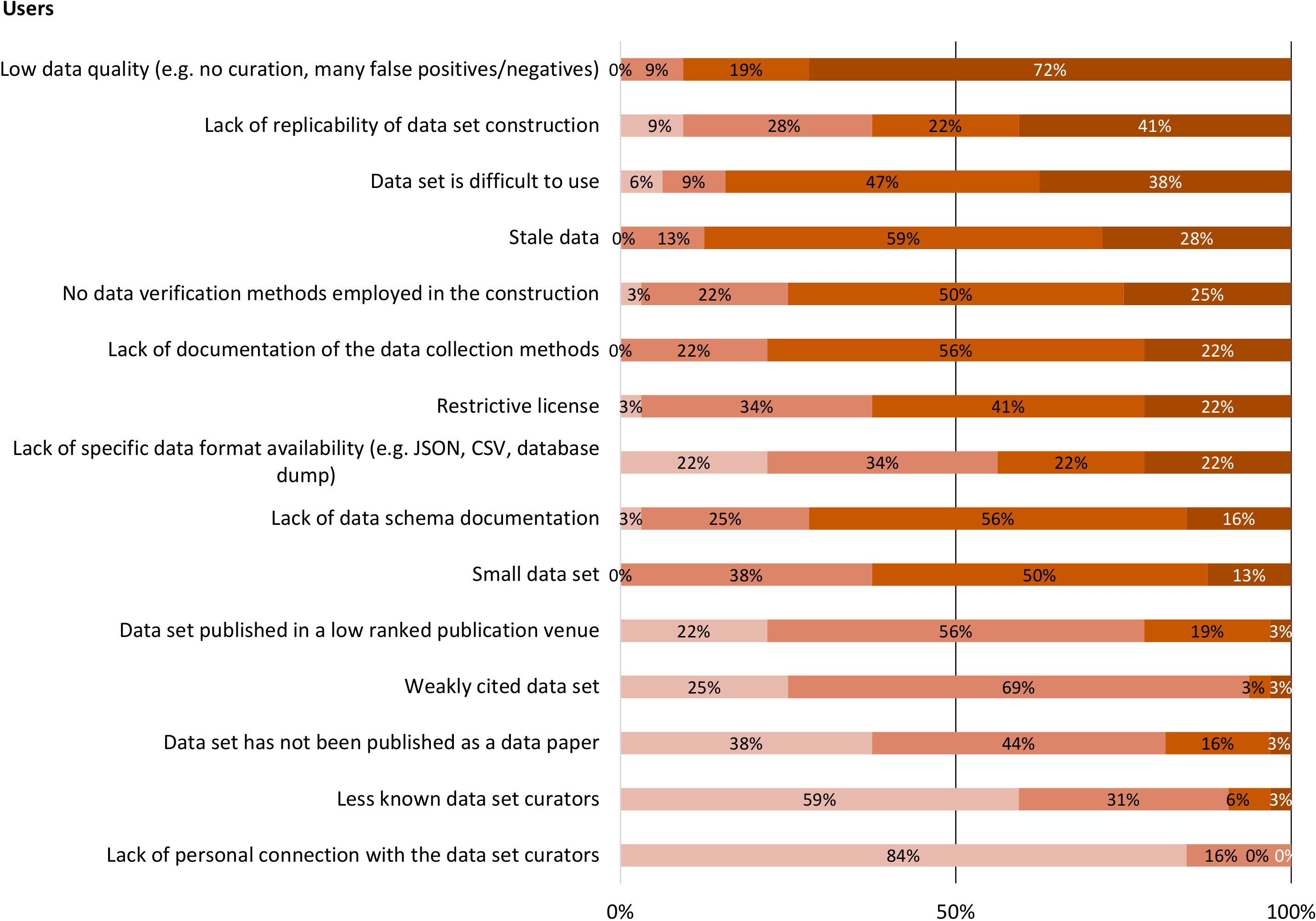}
\caption{Discouraging characteristics in data set selection
for data paper authors and users (top), and users isolated (bottom).
Respondents assessed a set of 15 predefined characteristics
on a four-level Likert scale.
Among the most discouraging
characteristics, low data quality stands out.}
\label{fig:preventing-characteristics}
\end{figure}

For this question,
all 108 respondents evaluated the degree to which
a set comprising of the negations
of the attributes presented in the previous question
discourage them from using a data set.
The evaluation was again done on a four-level Likert scale
and the results are presented
in Figure~\ref{fig:preventing-characteristics}.
Again,
the attributes are sorted downwards in descending order
of the \emph{Extremely} level.
Among the most discouraging characteristics,
we discerned
\emph{low data quality},
\emph{difficulty of data set use},
\emph{lack of documentation of the data collection methods},
\emph{lack of replicability of data set construction},
\emph{restrictive license},
and \emph{lack of data schema documentation}.
Less discouraging characteristics included
\emph{less known data set curators},
\emph{lack of personal connection with the data set curators},
and \emph{data set not having been published as a data paper}.
Isolating data paper users from authors
in Figure~\ref{fig:preventing-characteristics},
we notice the same characteristics in the last places
and in the first position,
with the intermediate attributes varying.
Users placed
\emph{lack of replicability of data set construction},
\emph{stale data},
and \emph{no data verification methods employed in the construction}
higher than authors and users combined.

Except for the above characteristics,
respondents further recognized through the related open-ended question,
the discouragement derived from
troubled access such as download issues,
difficult access,
and tool incompatibility/dependency,
data isolation,
lack of feature variety, novelty, support by the curators,
or extensibility,
and the limited scope of a data set.

\textbf{Q4.5: What direction should data papers follow in the future?}

\emph{Future data papers.}
According to the 108 responses,
future data papers could draw from
artificial intelligence and machine learning,
alternative and evolving software engineering,
logs analysis (e.g. build logs, test failure logs),
continuous integration and DevOps (particularly operations of DevOps),
collaborative software development,
code metrics and analyses,
cross-disciplinary or domain specific processes.
Furthermore,
the respondents would appreciate shared data regarding
health, fitness and performance,
repository duplication and code cloning,
human-centered and human-assessed data,
concerning new technologies,
and referring to
security,
social media,
online computing courses,
programming competitions,
video material and video games.
Data from
software services,
industrial ecosystems,
grey literature, 
databases,
Internet of things firmware,
voluntarily collected data, and
data on hyper-parameter optimization are also desired.
Lastly,
some participants suggested
conducting systematic literature reviews on published data sets,
and producing metadata after curating them.

\emph{Future data sources.}
As far as data sources are concerned,
many similar responses with the previous set of answers were observed.
Authors and users of data papers would like to see data extracted from
sources of entertainment, such as music and video streaming,
smart devices,
surveys,
sources with documented and valid data,
as well as data from the sectors of
education,
health,
energy,
defense and security,
manufacturing and retail,
blockchain,
finance,
and autonomous driving.
Moreover,
participants suggested exploiting
Alexa Rank,
execution logs,
code review systems,
integrated development environments,
activity sensors,
domain specific data sources,
and sources complementary to software repositories.
Industrial cloud systems,
safety critical software systems,
software industries,
human resources,
and industries with security breaches are also proposed
as prospective sources of data.

Separating data paper users from authors,
no variation was observed in their responses, since
the majority of the above ideas regarding future data papers and data sources
were also included in their answers.

\textbf{Feedback}

\emph{Overall.}
95\% (103) of the respondents assessed the completeness of the questionnaire
as \emph{sufficient}, whereas the remaining 5\% (5) reviewed it as \emph{weak}.
Overall,
the survey was characterized as good and concise,
with interesting aspects and sufficient maps to the objective questions.
Still, a respondent was not sure whether the open-ended feedback question
was enough for assessing completeness.

\emph{Objective questions.}
Q1 was characterized as answerable,
as opposed to Q2 which was characterized as ambiguous.
A few participants separated the process of data retrieval as part of another research paper requiring different effort
from the data paper composition.
This could be the reason we observe a significant wide spread in the responses.
Q3 and Q4 were considered as mirroring questions by some respondents.
However, the results of both the pilot and the final study partially contradict
this opinion due to the asymmetry observed in the answers
on an individual level.
For instance,
although \emph{ease of use} is ranked first in Figure~\ref{fig:useful-characteristics}
with 63\% of respondents considering it \emph{very important},
it is placed second in Figure~\ref{fig:preventing-characteristics}
with 46\% of respondents considering it \emph{extremely} discouraging.
Similarly,
\emph{high data set curator reputation} was evaluated as \emph{important}
by 48\% of participants,
but only 8\% evaluated it as \emph{significantly} discouraging.
We also observe that
\emph{data freshness} is placed considerably higher
than the equivalent \emph{stale data}.
Regarding Q5, a few respondents considered it enjoyable
as opposed to some others who characterized it as ambiguous
and unanswerable.
Moreover,
one participant remarked the lack of
neutral option in the Likert scale questions.

\emph{General comments.}
Apart from comments on the questionnaire,
some participants highlighted the significant effort required
to produce a data paper,
especially highly citable ones.
According to them,
these are a combination of
good tooling architecture,
documentation,
novelty,
generality,
reproducibility,
and clarity.
Nevertheless,
others expressed their concerns
regarding data paper practical issues and troubled data sharing guidelines, such as the General Data Protection Regulation.
Finally, an additional question was suggested
concerning the number of existing data sets
that researchers have used in their research.

\section{Discussion and Implications}
\label{sec:discussion}

As evidenced by the large increase in the published
data papers since the \msr\ data showcase track was formalized in 2013,
it is apparent that the track has catalyzed the publication
of data papers.
With data papers being more than 15\% of the \msr\ publications in 2019,
it is clear that the \msr\ data showcase track has spurred a new type of publication,
yielding each year a notable number of studies.
More generally, the data showcase track's success in driving the publication
of data papers indicates that a suitably themed conference track can in
some cases drive research toward a given direction.

The categories of data papers (Table~\ref{tab:data-paper-clusters})
span equally product and process, but product-oriented papers outnumber
the process ones.
This can be explained by the preponderance of publicly available
product data, which are associated with open source software projects,
over process data, which are more difficult to come by.
Although past experience with calling
for the publication of particular data types has not been encouraging~\citep{Wal98}, many years have passed since then and it might be worth to try focusing the \msr\ call for data papers on specific topics each year,
with an emphasis on software processes,
in order to overcome the previous bias.

The studies that strongly cite data papers span the \textsc{swebok} knowledge
areas fairly unequally.
It seems that software maintenance and engineering management can be profitably studied
using materials from \msr\ data papers, but
software requirements, economics, and engineering models and methods less so.
Given the, by definition, primary importance of all \textsc{swebok} areas,
it would seem that the \msr\ data showcase track chairs
could promote studies associated with the less covered areas
by adjusting the track's call for papers to specifically invite data
sets targeting them.
We acknowledge, however, that for certain \textsc{swebok} areas,
such as software economics, the release of data sets is hard
due to the often proprietary nature of the corresponding data,
while in others, such as software requirements,
there is an established tradition to publish data sets
together with research papers~\citep{ZSMA17}.
Nevertheless, data sets for underrepresented \textsc{swebok} areas
might have lasting impact in their subfield
despite being less popular.

\begin{implication}
The \msr\ data showcase track chairs could target the call
for data papers on process-oriented topics and less covered \textsc{swebok} areas,
to possibly improve their footprint.
\end{implication}

Although one might expect that a data paper is typically cited
mainly when it is actually used, our findings do not support this assertion.
We manually identified 440 strong citations;
far fewer than half of the
1169 total citations that were made to data papers according to our results.
This demonstrates that citations to any kind of published studies
(including data research) can be made for a variety of reasons.
According to the manual analysis of the weak citations,
the most prominent reasons are mentioning the work as an example,
attributing a work's statement, and using the work's methods.
Be that as it may, based on the difference between the citations
to data papers and to other studies, there seems to be room
for improving the data papers' use.

\begin{implication}
The actual use of data papers could potentially be increased
through the promotion of open science initiatives
by journal editors and conference program committees,
such as the \textsc{acm} Artifact Review and Badging policy~\citep{Boi16}.
\end{implication}

With each data paper strongly cited on average 5.4 times, it appears that
data papers are in general useful for conducting other empirical studies.
Many of these studies are published in top-notch venues
(see Table~\ref{tab:venues}),
indicating the high quality of studies that can be performed
through data papers.
On the other hand, at least for \msr\ papers that cite data
papers,
their basis on published empirical data does not seem to increase
their impact in terms of citations to them
(see last column of Table~\ref{tab:citation-metrics}).

Regarding impact,
the number of strong citations to data papers is constantly rising
(Figure~\ref{fig:yearly-papers}), indicating that the concept of data papers
has long-term value.
The enduring usefulness of specific papers
is also apparent by looking at the timeline of strong citations
to specific \msr\ data showcase papers over a five-year period
(Figure~\ref{fig:five-year-citations}).
While the majority of papers indicate a short shelf life,
the trend of the most strongly cited data papers retaining their
citation number, or obtaining even more strong citations, is yet another
manifestation of the Matthew effect in science~\citep{Mer68}.
However, this results in a constant need for new data papers,
which was also expressed by some respondents of the survey
stating that fields with ongoing changes
result in a short shelf life for the corresponding data sets.

\begin{implication}
The short shelf life of data sets implies
a need for a constant stream of new data papers.
\end{implication}

\begin{figure*}[t]
\centering
\includegraphics[width=0.49\textwidth]{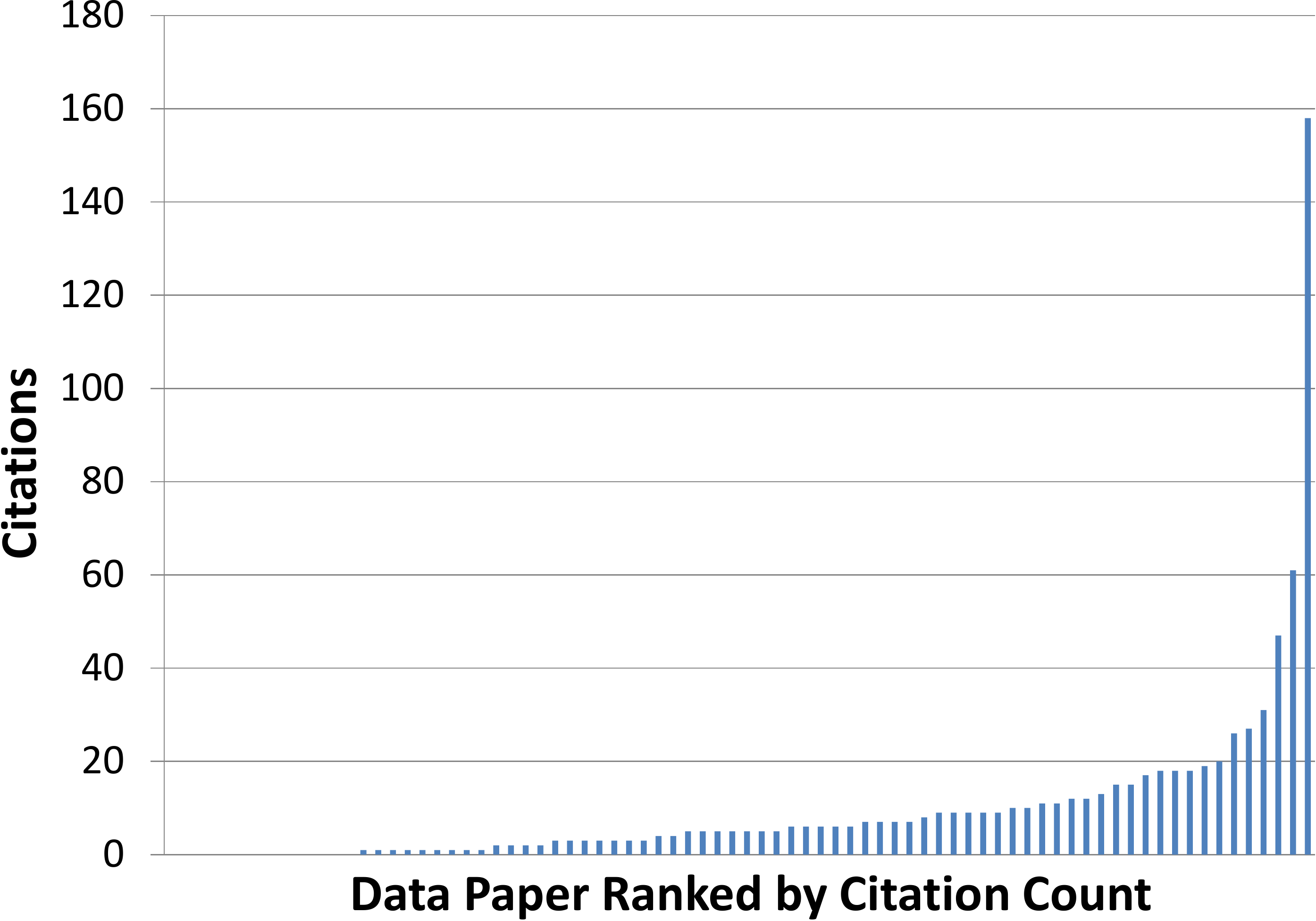}
\includegraphics[width=0.49\textwidth]{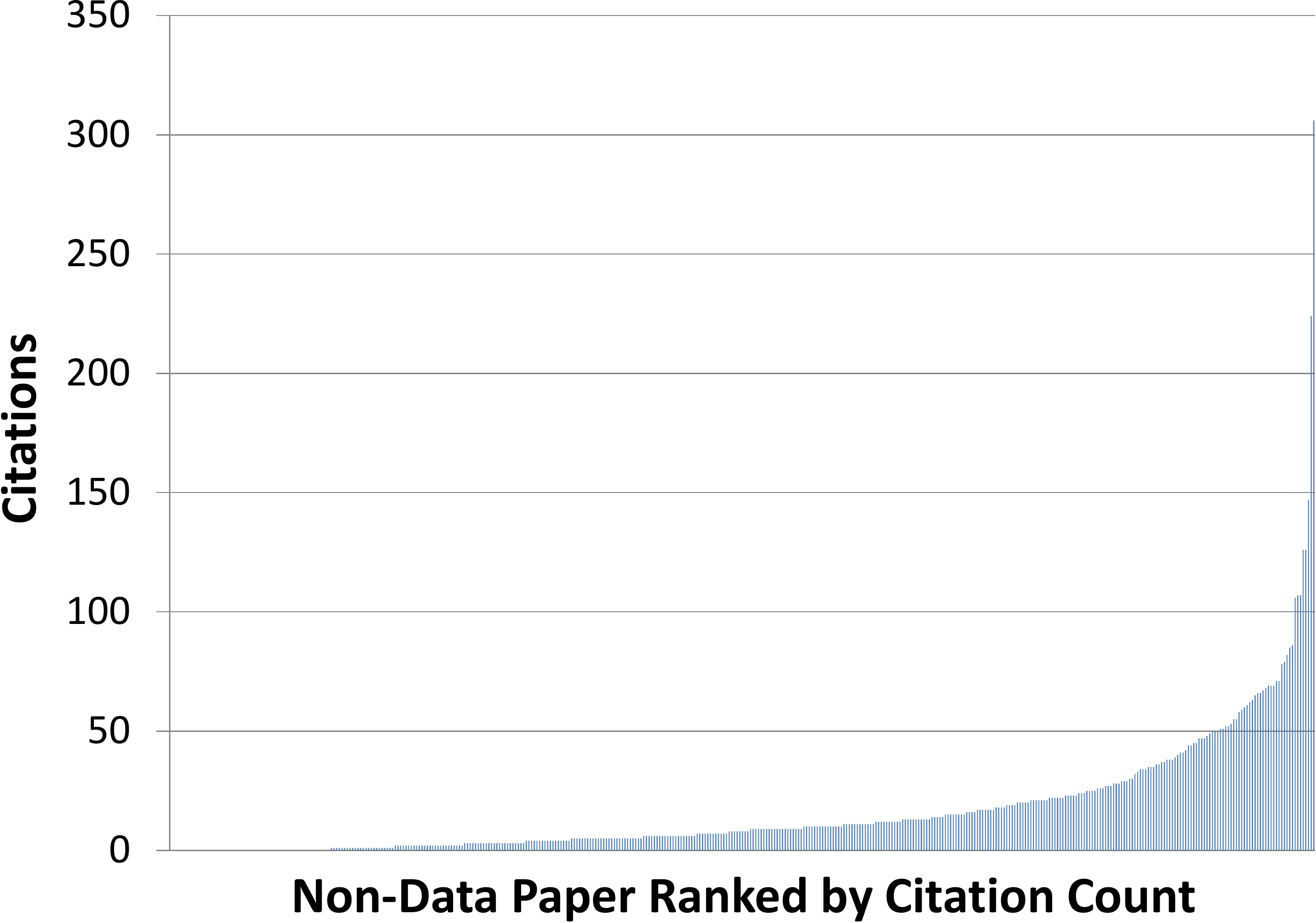}
\caption{Distribution in the number of citations to MSR data papers (left) and MSR non-data papers (right). The similar shape of the two
distributions indicates that the reason for the lower citation
count of MSR data papers is the overall lower number of
citations to each data paper compared to the citations to
each non-data paper.}
\label{fig:citation-distribution}
\end{figure*}

Yet,
surprisingly for an artifact whose main purpose is for others to build on,
data papers are cited less than other \msr\ papers.
One might think that this is due to the 28 out of
81 (35\%) of the data papers that are never used.
The citation's distribution long tail --- just 9\% of the data papers are
strongly cited by 67\% of all citing studies --- could be another reason.
However, by comparing the distribution of citations to data papers
(according to Scopus) with that of citations to non-data papers
(Figure~\ref{fig:citation-distribution}), we see that the two
distributions are similar in shape.
It is apparent that the reason for the lower citation count of
\msr\ data papers is the overall lower number of citations to each data paper
compared to the citations to each non-data paper.

There are three reasons that could explain this phenomenon.
First, data papers may not publish data
that are actually useful for conducting other studies.
Our author survey suggests that high data quality,
ease of use, and data freshness are the most valuable characteristics
in data set selection.
To address this problem,
authors of new data papers can draw from these findings
and ensure the satisfaction of the particular criteria in their work.
In addition,
the \msr\ program committee could adopt more
stringent criteria for accepting data papers,
though this will certainly lead to a decline in the number of accepted papers,
and there is no guarantee
that a more selective track will still select the papers 
that will be most frequently cited.
The track's toughening of data sharing can be counterbalanced
again by promoting open science initiatives.

\begin{implication}
Program committees could consider adopting
more stringent criteria for accepting data papers,
to potentially improve their usage.
\end{implication}

Second, existing data papers may not satisfy the needs and interests
of software researchers.
The responses propose exploiting data related to topics
such as artificial intelligence and machine learning,
collaborative software development,
health, fitness and performance,
online computing courses, video material and video games.
Suggested future data sources include the sectors of
education, health, energy, manufacturing, and autonomous driving,
entertainment, and smart devices.

\begin{implication}
Prospective authors of data papers can exploit
the survey's insights to produce quality work that will meet
the community's expectations, needs, and interests.
\end{implication}

Third, researchers may be reluctant to work with data coming
outside their organization --- also
known as the \emph{not invented here syndrome}~\citep{PD15},
or fear that working with publicly available data is less likely
to yield original results.
Although respondents of the survey reported that personal connections
with data paper authors play little role in selecting data papers,
the high number of papers used by their authors (Figure~\ref{fig:self-citations})
contradicts this.
This practice suggests the possibility of adopting a workflow similar
to that of pre-registered studies~\citep{HI18}:
publishing a data paper and then employing
it for empirical software engineering research.
Such a workflow may further strengthen the safeguards
promoted by pre-registered studies
against p-hacking and publication bias~\citep{Kup18}.
In addition, encouraging the advance publication of a study's
data would level the playing field between the scientists with access
to rich empirical data and those without.

\begin{implication}
Methodology researchers,
conference program committees,
and journal editorial boards could examine the
opportunities and implications associated with a research paradigm
where the data employed in empirical studies are published before
the studies that analyze them.
\end{implication}

Looking at the most used data paper, GHTorrent~\citep{Gou13}, 
we observe that it is characterized by the majority of the attributes
considered useful by the respondents~(Figure~\ref{fig:useful-characteristics}).
Particularly,
one highlighted that GHTorrent's updatability through its available source code
``lends credence to the construct validity of the data set,
since the instrument used to curate it is open for review''.
Adding to that the continuous human effort and attention
by the curators for its regular maintenance,
through daily and bi-monthly database dumps accessible
from its web site,\footnote{\url{https://ghtorrent.org}}
addressing users' suggestions and bug reports
submitted to the GitHub
project,\footnote{\url{https://github.com/ghtorrent/ghtorrent.org}}
this seems to create a self-reinforcing feedback loop
between the curators' efforts and the continuing citations to it.

\begin{implication}
Self-reinforcing feedback loops
could affect positively a data set's citations over time,
starting from the curators' regular data set updates, maintenance,
and support.
\end{implication}

Overall,
data paper authors tend to publish work with data they have gathered
in the context of other research.
Still, at the same time they seem motivated
to benefit the community with new data,
and they consider that a worthwhile way of doing so is
by publishing data papers.
Encouragingly, our author survey paints a picture of an
open and meritocratic community,
with authors failing to agree that drafting a data paper is an easy way to pad
a {\sc cv} with more publications.
Moreover, they seem notable supporters of open science,
through various open-ended responses
where they expressed their desire for furthering open science goals.
Hence, the suggestion regarding the promotion of open science
initiatives is enhanced through the particular observation.

\section{Threats to Validity}
\label{sec:validity}

The study's external validity in terms of generalizability, obviously
suffers by studying only data papers that have been published within
the framework of the \msr\ conference and ignoring venues such
as the \textsc{promise} conference --- consider e.g. the work by \citet{FTLS18}
---
or the \emph{Empirical Software Engineering} ---
e.g. the paper by \citet{Squ18}.
However, studying the \msr\ conference in isolation
allowed us to analyze the effect of establishing the \msr\ data showcase track,
and to compare citation counts among different groups of papers
(Section~\ref{sec:citation-analysis}),
without the bias associated with a paper's publication venue.
Furthermore,
external threats are also related to our ability to generalize
the author survey results.
Again, the sample selection only within the \msr\ boundaries
prevents us from generalizing our conclusions to other venues.
Still, it allowed meaningful insights to be derived,
which could be enhanced and generalized
through replication of the study to other venues.

The major threats to the study's internal validity stem from the steps
during which we followed manual processes involving subjective judgment:
the selection of data papers before the showcase track was introduced,
the filtering of studies that actually use data papers,
the analysis of the weak citations,
the clustering of data papers,
the classification of studies using data papers, and
the pair coding of open-ended responses.
Especially the clustering of data papers introduced
in Table~\ref{tab:data-paper-clusters}
holds another serious threat associated with the establishment
of the clusters themselves.
As elaborated in Section~\ref{sec:data-paper-methods},
clusters resulted from a conceptual analysis of the corresponding
data studies.
The risk stemming from the pair coding process is related
to the loss of accuracy of the original response due to an
increased level of categorization.
The trustworthiness of the processes of clustering, classification,
and pair coding were enhanced through the use of multiple raters
and coders,
and by grounding them on established research methods.
However, we acknowledge that validity risks derived
from manual processes requiring human judgment
cannot be completely eliminated~\citep{PVK15}.
Another threat related to the survey responses is
social desirability bias~\citep{Fur86}
(i.e., a respondent's possible tendency to appear in a positive light,
such as by showing they are fair or rational).
Particularly,
the answers presented in Figure~\ref{fig:motivational-factors}
may lack some truthfulness.
For instance, one should not over-interpret
that few respondents consider the \msr\ data showcase track
\emph{a straightforward way to publish in the MSR conference}.
To mitigate this bias, participants were informed
that responses would be anonymous.
Question-order effect~\citep{Sig81}
(e.g. one question may have provided context for the next one)
may have led respondents to a specific answer,
especially in the answers presented in Figures~\ref{fig:useful-characteristics}
and \ref{fig:preventing-characteristics}.
One approach to mitigate this bias could have been
randomizing the order of questions.
In our case,
we decided to order the questions in a convenient manner
for respondents to easily recall and understand the context
of the questions asked.

\section{Conclusions}
\label{sec:conclusions}

The \msr\ data showcase track has been successful in encouraging the
publication of data papers.
Data papers are generally used by other empirical studies,
though not as much as one might expect or hope for.
The gatekeepers of science, such as journal editors and program committees,
can address this by setting a higher bar for the publication of data papers, by encouraging their constant stream and use, and by promoting open science initiatives.
An additional policy to improve the use and impact of data papers
might be to provide incentives for researchers to enrich existing
collections of data instead of reproducing similar data sets from scratch.
Such incentives could involve awarding a most influential data paper award
or inviting papers where researchers describe how they expanded
upon a data track study.

\begin{acknowledgements}
Panos Louridas provided insightful comments on this manuscript.
Furthermore,
Georgios Gousios's suggestions regarding the refinement of the questionnaire
were crucial for the survey attainment.
This work has received funding from:
the European Union's Horizon 2020 research
and innovation programme under grant agreement No 825328;
the {\sc gsrt} 2016--2017 Research Support (EP-2844-01); and
the Research Centre of the Athens University of Economics and
Business, under the Original Scientific Publications framework 2019.
\end{acknowledgements}

\section*{Conflict of Interest}

The authors declare that they have no conflict of interest.

\bibliographystyle{spbasic}
\bibliography{data_papers,strong_citations,KKDS19}

\end{document}

%% file: data-papers-by-year.tex
2005 & \citep{SSHP05} \citep{MLRW05} \citep{CHC05}  \\
2006 & \citep{KZKH06}  \\
2010 & \citep{NZ10}  \\
2012 & \citep{KFHE12}  \\
2013 & \citep{BLPH13} \citep{DHPK13} \citep{GCM13} \citep{VSM13} \citep{WJS13} \citep{JHSA13} \citep{MK13} \citep{Squ13a} \citep{MBR13} \citep{BWYS13} \citep{Squ13b} \citep{LPD13} \citep{Gou13} \citep{RVV13} \citep{HKYC13}  \\
2014 & \citep{KL14} \citep{SSOL14} \citep{GZ14a} \citep{MHK14} \citep{PC14} \citep{ZH14} \citep{RASV14} \citep{LRS14} \citep{BAKO14a} \citep{FHMF14} \citep{GVSZ14a} \citep{WDMD14} \citep{FTC14} \citep{MKLG14} \citep{BP14}  \\
2015 & \citep{VSF15a} \citep{SB15} \citep{OKYY15} \citep{KMMR15} \citep{GAH15a} \citep{ASDW15} \citep{Spi15} \citep{WY15} \citep{MBSG15} \citep{BLSS15} \citep{KMLG15} \citep{PNTB15} \citep{PML15} \citep{Zac15} \citep{HMMB15a} \citep{GRI15}  \\
2016 & \citep{PANM16a} \citep{ABKL16} \citep{Squ16} \citep{YKYI16} \citep{ANNN16} \citep{ZZM16} \citep{OMDT16}  \\
2017 & \citep{NMS17} \citep{AHMR17} \citep{ZLRC17} \citep{RHHC17} \citep{MK17} \citep{SBM17} \citep{YAKG17}  \\
2018 & \citep{MAL18a} \citep{YLYW18} \citep{NCL18} \citep{GMPP18a} \citep{XZ18} \citep{SLLY18} \citep{PKHH18} \citep{YPKG18a} \citep{Spi18} \citep{GYJL18a} \citep{CPDT18} \citep{ML18a} \citep{SZC18} \citep{GMS18} \citep{ECS18}  \\

%% file: citation-metrics.tex
N	&	78	&	78	&	429	&	49	\\
Min	&	0	&	0	&	0	&	0	\\
Max	&	158	&	107	&	306	&	147	\\
Median	&	5	&	10	&	8	&	10	\\
Avg	&	9.8	&	16.9	&	17.0	&	15.7	\\
Stddev	&	19.9	&	21.7	&	27.7	&	25.3	\\

%% file: venues.tex
MSR & 52 & 13.8 \\
ICSE & 24 & 6.4 \\
CoRR & 21 & 5.6 \\
ICSME & 16 & 4.2 \\
SANER & 14 & 3.7 \\
EmpSE & 13 & 3.4 \\
ESEM & 6 & 1.6 \\
IEEE TSE & 6 & 1.6 \\
Other conference & 176 & 46.7 \\
Other journal & 49 & 13.0 \\